\documentclass[12pt,showkeys]{article}
\usepackage{amsfonts}

\newcommand{\be}{\begin{equation}}
\newcommand{\ee}{\end{equation}}
\newcommand{\ba}{\begin{eqnarray}}
\newcommand{\ea}{\end{eqnarray}}
\newcommand{\pl}{\left\{}
\newcommand{\pr}{\right\}}
\newcommand{\al}{\left|}
\newcommand{\ar}{\right|}
\newcommand{\rr}{\right)}
\newcommand{\rl}{\left(}

\begin{document}

\begin{titlepage}
\rightline{{CPHT RR 002.0104}}
\rightline{{LPT-ORSAY 04-11}}
\rightline{hep-th/0401201}

\vskip 2cm
\centerline{\large \bf Non-tachyonic Scherk-Schwarz compactifications,}
\centerline{\bf \large cosmology and moduli stabilization}
\vskip 1cm
\centerline{E. Dudas${}^{\dagger,\star}$ and C. Timirgaziu${}^{\star}$}
\vskip 0.5cm
\centerline{\it ${}^\dagger$ Centre de Physique
  Th{\'e}orique\footnote{Unit{\'e} mixte du CNRS et de l'EP, UMR 7644.} ,
Ecole Polytechnique, F-91128 Palaiseau}
\vskip 0.3cm
\centerline{\it ${}^\star$ Laboratoire de Physique 
Th{\'e}orique
\footnote{Unit{\'e} Mixte de Recherche du CNRS (UMR 8627).}}
\centerline{\it Universit{\'e} de Paris-Sud, B{\^a}t. 210, F-91405 Orsay Cedex}

\vskip  1.0cm
\begin{abstract}

It is well-known that Scherk-Schwarz compactifications in string theory have a tachyon in
the closed string spectrum appearing for a critical value of a compact radius. The
tachyon can be removed by an appropriate orientifold projection in type
II strings, giving rise to tachyon-free compactifications. We present explicit examples
of this type in various dimensions, including six and four-dimensional chiral
examples, with softly broken supersymmetry
in the closed sector and non--BPS configurations in the open sector. 
 These vacua are interesting frameworks for studying various cosmological issues.   
We discuss four-dimensional cosmological solutions and moduli
stabilization triggered by nonperturbative effects like gaugino
condensation on D-branes and fluxes. 

\end{abstract}
\end{titlepage}

\section{Introduction}

Supersymmetry breaking in string theory is a still open and central
question for string phenomenology and cosmology. At the perturbative 
string level there are basically three known ways of breaking supersymmetry~:
Scherk-Schwarz type compactifications \cite{rohm}, breaking induced by
internal magnetic fields (T-dual to branes intersecting at angles)
\cite{bachas} and breaking by non-BPS configurations \cite{bsb}.  
There are also completely nonsupersymmetric heterotic or type O string
constructions \cite{nonsusy,augusto2}. 
 
Scherk-Schwarz string compactifications have generically a
tachyon-like field in their spectrum that appears for a critical value
of a compact radius
$R^2 = 2 \alpha'$ \cite{rohm}. Orientifolds \cite{augusto1} (see
\cite{reviews} for reviews on orientifold constructions and extensive
references) of such vacua were
constructed in \cite{bd} and \cite{ads1}. It was then shown in
\cite{dm3}, \cite{dmt1} that the lowest mass would-be tachyon field can
be eliminated by using an orientifold projection similar to the one
used in type O strings \cite{augusto2}.   
In two more recent papers \cite{dmt1,dmt2}, we studied classical solutions
of these models and found nonsingular 9D solutions in the compact space,
related by a change of coordinates to the supersymmetric solution with
(constant velocity) moving spacetime boundaries. A compactification to
four dimensions was also performed. As a result, an expanding FRW universe was
found, together with a natural way of producing three large spacetime dimensions
and a much slower time evolution of the compact space.

The goal of this paper is to generalize these results to
lower dimensions, construct vacua which have the necessary features for finding 
later on realistic examples, in particular chiral
fermions and start studying nonperturbative issues like moduli
stabilization. The stabilization will change the cosmological
solutions into Minkowski or de Sitter solutions, depending on details
of moduli stabilization.    

The paper is organized as follows. Section 2 describes the basic
brane-orientifold plane system appearing in all the non-tachyonic
constructions of our paper and its basic features. Section 3 presents
an eight dimensional example where the two-dimensional compact space
is a tilted torus, and a chiral six dimensional example. Section 4  
presents four-dimensional examples based on $Z_2 \times Z_2$
orbifolds, including a chiral model based on a model with discrete torsion.  

Section 5 represents a first excursion into the issue
of (Kahler) moduli stabilization triggered by nonperturbatively induced brane
potentials, like for example gaugino condensation, and/or 
fluxes. We show that nonperturbative dynamics can stop the moving
brane and stabilize the modulus describing the distance
between the boundaries and also the other volume Kahler modulus, in
analogy with the phenomenological analysis performed in \cite{gw} in a
different context. In an example with nonpertubative gaugino
condensation, naturally realized in our string examples,
stabilization is possible only with gaugino condensation combined with
constant terms in the potential in a positive perfect square, in close
analogy with the four-dimensional heterotic examples of gaugino
condensation. We find also a no-scale structure for the dilaton, which
is therefore not stabilized by the brane dynamics we are
discussing. In order to stabilise the dilaton we could add in addition
a combination of NS-NS and RR fluxes, following \cite{gkp}.
The resulting spacetime can be Minkowski or de Sitter, depending on the 
parameters entering the stabilization mechanism. 
 
The Appendix provides our conventions and useful formulae for $Z_2
\times Z_2$ characters used in the text.



\section{ The Dp-brane / ${\overline{Op}}_{-}$-plane system in string theory }

Time-dependence in string theory relies until now on some simple and
computationally tractable examples \cite{epkyrotic,lms}. The best known example is the
brane-antibrane pair. This system breaks completely supersymmetry and
the open string spectrum contains a tachyon stretched between the brane
and the antibrane, reflecting the attraction between them. The open
string tachyon condensation and, more recently, the detailed time
evolution triggered by the tachyon field was subject to intense study
starting from the seminal papers of A.Sen \cite{sen}. 

An interesting but much less studied system is the Dp-brane /
${\overline{Op}}_{-}$-plane system. In our terminology, the
${\overline{Op}}_{-}$-plane is a (non-dynamical) anti-orientifold plane of $p+1$
world-volume dimensions having the same quantum numbers as a
coincident superposition of $2^{p-4}$ antibranes ${\overline{Dp}}$. 
The branes and the antiorientifold planes we consider attract each
other, in close analogy with the brane-antibrane system. The
orientifold planes are non-dynamical objects, however, and this implies that
they cannot annihilate with the branes. This reflects, in particular,
in the fact that there is no tachyon stretched in the system we
consider. Their mutual attraction, however,
clearly generates a time-dependence which is one of the main motivations
for the string constructions in the present
paper. The tension and RR charge of a system of $n$ $Dp$ branes and an
${\overline{Op}}_{-}$ plane are given by
\be
\mathcal{T} = (n+2^{p-4}) \mathcal{T}_p \quad , \quad q = (n-2^{p-4})
\mathcal{T}_p \ , \label{s1}  
\ee
where  $\mathcal{T}_p$ is the tension of an elementary Dp brane,
and are such that $\mathcal{T} > q$. Let us now place ourselves in a
reference frame in which the system moves with a constant velocity
$v$. In that frame, the effective tension and RR charge are 
$(T_{\rm eff} = T \sqrt{1-v^2} , q)$.  If we observe the non-BPS system
in a frame where $T_{\rm eff} =q$, then it will look as the BPS system  
n Dp-$Op_{+}$, where the (usual) $Op_{+}$ plane has the tension and RR charge 
$(-2^{p-4},-2^{p-4})$. In particular it will consistently couple to the
static and supersymmetric geometry corresponding to an appropriate BPS system. 
Therefore the system we are considering, 
$n \ Dp - {\overline{Op}}_{-}$, moving with a constant velocity $v = th \ \xi$, where
\be
ch \ \xi = {n + 2^{p-4} \over n - 2^{p-4}} \ , \label{s2} 
\ee
behaves similarly to the BPS configuration $n \ Dp - {Op}_{+}$. 

The explicit string examples we provide in Sections 3 and 4 contain
actually, in addition to the system mentioned above, additional branes
and conventional orientifold planes $O_{+}$. A further interesting property
of this class of string vacua is that the $O_{+} - {\overline O}_{-}$ 
system, even if of orientifold-antiorientifold type, has the property of
eliminating the closed string tachyon present in the string vacua
before orientifolding. More precisely, as shown in the 9d example
constructed in \cite{dmt1} and in the lower dimensional examples in Section 3
and 4, $O_{+} - {\overline O}_{-}$ pairs appear in Scherk-Schwarz
compactifications (with a peculiar orientifold projection), which have
in the closed string spectrum a would-be tachyon \cite{rohm}.
Standard string techniques show that $O_{+} - {\overline{Op}}_{-}$
interaction is described by
(for the definition of the characters, see the appendix)
\ba
&& {\rm loop \ channel} \qquad - \ (O_8-C_8) \ , \nonumber \\
&& {\rm tree-level \ channel} \qquad - \ (V_8+S_8) \ , \label{s3}
\ea   
where, as usual, in the Klein bottle amplitudes, $(O,V)$ describe the NS-NS
sector (a scalar tachyon, graviton and antisymmetric tensor), whereas
$(S,C)$ describe the RR sector. These amplitudes are similar to the
brane-antibrane one, but their interpretation is quite different, since
they just symmetrize (or antisymmetrize) states already existent in the
torus amplitude. In particular, the scalar closed string tachyon is antisymmetrized and
is therefore eliminated from the spectrum. 

More precisely, as shown in \cite{dmt1} and discussed in more detail in
the compactified models in Sections 3 and 4, these exotic orientifold
configurations are generated by the orientifold projection $\Omega'= \Omega \Pi_{y}
(-1)^{f_L}$, where $(-1)^{f_L}$ is the left 
world-sheet fermion number\footnote{A similar way of eliminating the closed
string tachyon was first proposed in the context of Type O orientifolds
by Sagnotti \cite{augusto2}. Its implementation in Scherk-Schwarz compactifications 
was proposed in \cite{dm3}.} and $y$ is the coordinate used to break
supersymmetry. The lowest mass tachyonic states of zero Kaluza-Klein
momentum $m=0$ and winding number $w=2n+1=\pm 1$ are odd under it and therefore
eliminated from the spectrum. 
The lowest mass scalar states in the spectrum are the antisymmetrized combinations
\ba
&& | T_1 > = |m=+1,w=1> - \  |m=-1,w=1> \ , \nonumber \\ 
&& | T_2 > = |m=+1,w=-1> - \  |m=-1,w=-1> \ . \label{s4} 
 \ \label{t1}
\ea
This orientifold projection removes the closed tachyon for
any radius $R$ since the mass of the states (\ref{s4}) 
\be
M_T^2 = -{2 \over \alpha'} + {1 \over R^2} + {R^2 \over \alpha'^2} \ = \ 
({1 \over R}- {R \over \alpha'})^2 \  \label{s5}
\ee
is positive and becomes zero at the self-dual value of the compact radius. 
It is easily checked that in Scherk-Schwarz compactifications
$\Omega'$ squares to one because $(-1)^{f_L+\bar f_R}=1$ in all
sectors. The involution $ \Omega'$ generates $Op_{+}$-planes at $y=0$
and  anti $O8_{-}$-planes at $y=\pi R$. 

These features are present in all explicit string theory examples and
will become more transparent in the vacua constructed in Sections 4 and 5. 
 
\section{String vacua in various dimensions }

In this section  we present explicit string constructions realizing the 
Dp-brane / ${\overline{Op}}_{-}$-plane system introduced in Section 2. All of the explicit constructions start from a freely-acting
orbifold $g$ in the type II string, containing the spacetime fermion
number $(-1)^F$. After a radius redefinition, 
the orbifold $g$ becomes a periodic identification $y = y + 2 \pi R$
accompanied by the spacetime fermion number operation, imposing
different boundary conditions for bosons and fermions and breaking
therefore supersymmetry. The orientifold operations $\Omega' = \Omega
\ \Pi_y \ (-1)^{f_L}$ and $\Omega' \ g$ create O-planes of two
different types. The fixed plane of $\Omega'$ sits at the origin $y=0$
and is a standard $O_{+}$ plane, whereas the fixed plane of $\Omega' \
g$ sits at $y = \pi R$ and is an antiorientifold plane, due to the
action of $(-1)^F$. More precisely, it is an  ${\overline O}_{-}$ plane
due to the simultaneous action of $(-1)^F$ and  $(-1)^{f_L}$
operations. 
 
The simplest example of this type is nine-dimensional and 
was provided in \cite{dmt1}. A straightforward way of producing
lower-dimensional examples is by taking five additional dimensions to be
compact and performing T-dualities on the original nine-dimensional
example.
  There are several reasons, however, to search for new examples. The
first is that T-dualities on the 9d example will produce non-chiral
vacua in lower dimensions, whereas a realistic example asks for
chirality. In this respect, we construct six and four-dimensional chiral examples
by compactifying on orbifolds. The second reason is building
models with a rich spectrum of D-branes. Our
six and four-dimensional examples are based on $Z_2$ and $Z_2 \times Z_2$ orbifold
compactifications. 

\subsection{String vacua in 8 dimensions}

 The starting point for the eight-dimensional constructions is the $IIB$
superstring compactified on a $T^2$ torus, subject to the orbifold
identification $g=(-1)^F \ \delta$, where $F$ is the spacetime fermion
number and $\delta$ is the symmetric shift 
\be
\delta \ :  \ X_i=X_i+\pi R_i, \quad i=1,2 \ . \label{e1}
\ee
We introduce for
convenience the complex coordinate $z = X_1 + i X_2$. Then the original
circle identification  $X_i \equiv X_i+2 \pi R_i$ and the shift operation
$\delta$ are both contained in the identifications
\be
z \ = \ z + 1 \quad , \quad z \ = \ z + \tau \ , \label{e2} 
\ee  
where $\tau = (1/2) + i (R_2/2R_1)$. The shift identification has
thus the effect of turning the square torus into a tilted torus of complex
structure $\tau$. For later use we also define the Kahler modulus $\rho
= i (R_1 R_2 /2)$. 
  
The $g$ operation breaks completely supersymmetry and has no fixed points.
The corresponding torus amplitude reads:

\ba 
\mathcal{T}&=& 
\rl |V_8|^2+|S_8|^2\rr
(\Lambda_{2m_1,n_1} \Lambda_{2m_2,n_2}+\Lambda
_{2m_1+1,n_1} \Lambda_{2m_2+1,n_2} ) \nonumber\\
&+&\rl |O_8|^2+|C_8|^2
\rr(\Lambda_{2m_1,n_1+1/2} \Lambda_{2m_2,n_2+1/2} + 
\Lambda_{2m_1+1,n_1+1/2} \Lambda_{2m_2+1,n_2+1/2}) \nonumber\\
&-&\rl V_8\bar{S}_8+S_8\bar{V}_8 \rr(\Lambda_{2m_1,n_1}\Lambda_{2m_2+1,n_2}+
\Lambda_{2m_1+1,n_1}\Lambda_{2m_2,n_2}) \nonumber\\
&-&\rl
O_8\bar{C}_8+C_8\bar{O}_8\rr(\Lambda_{2m_1,n_1+1/2}
\Lambda_{2m_2+1,n_2+1/2}+\Lambda_{2m_1+1,n_1+1/2}\Lambda_{2m_2,n_2+1/2}) \
. \label{e3}
\ea
 
The most natural orientifold candidate for a freely-acting orbifold is
the one obtained with the operation containing one parity
$\Omega'=\Omega (-1)^{f_L} \Pi_2 $. The orientifold operation introduces
O8 planes, which require, for consistency, the presence of D8
branes. However, it turns out that the open string spectrum is completely
supersymmetric. Indeed, the Klein amplitude 
\be
\mathcal{K} = \frac{1}{2} \ (V_8-S_8) P_{2m_1} W_{n_2} \ \label{e4}
\ee
introduces standard $O8_{+}$ planes at $X_2 = 0 , \pi R_2$, asking for a
net number of 32 D8 branes. The open string amplitudes
\ba
\mathcal{A} &=& \frac{N^2}{2} \ (V_8-S_8) \ P_{m_1} \ ( W_{n_2} + W_{n_2+1/2} )
\ , \nonumber \\
\mathcal{M} &=&- \frac{N}{2} \
(\hat{V}_8-\hat{S}_8) \ [P_{m_1}W_{n_2}+(-1)^{m_1} P_{m_1} W_{n_2+1/2}] \ , \label{e5}
\ea
are completely blind to the supersymmetry breaking present in the closed sector. The
(maximal) gauge group, corresponding to putting all branes at the origin
$X_2=0$, is $SO(16)$. These amplitudes are (after a T-duality on $X_2$)
exactly the ones describing a discrete antisymmetric tensor background
in the Type I superstring \cite{augusto1}, a fact that explains the reduction
of the rank of the gauge group \cite{dmp}. The reason behind this simple
result is that geometric interpretation of O-planes / D-branes in this
model asks for a T-duality in $X_1$. After the T-duality, the      
complex structure and the Kahler modulus get interchanged $\tau
\leftrightarrow \rho$, such that the new ones become
\be
\tau' \ = \ i {R_1 \over 2 R_2'} \quad , \quad \rho' \ = \ {1 \over 2} +
i {R_1 R_2' \over 2} \ , \label{e6}
\ee
where $R_2'$ is the T-dual radius. In particular, by using the general
definition $\rho = B_{12} + i \sqrt{G}$, the presence of the quantized
antisymmetric tensor $B_{12} = 1/2$ is readily identified. 
This argument is equivalent to the known
\cite{ab} result that an orientifold action containing a parity
inversion in the presence of a nontrivial complex structure of the torus produces
a reduction of the gauge group. It is
interesting that the Klein and the open string amplitudes are exactly
the same as the Type I ones, whereas the closed spectrum has soft
supersymmetry, in analogy with the M-theory-type vacua constructed in
the last two references of \cite{ads1}. 
 
In order to construct a model having the pattern of O-planes put forward
in Section 2, we use a different orientifold projection. 
In order to have a more transparent geometrical interpretation, we first
rewrite the torus amplitude in a different (Scherk-Schwarz) basis, by a
trivial rescaling $R_i \rightarrow R_i /2$ of the internal radii. The
torus becomes:
\ba 
\mathcal{T}&=& 
\rl |V_8|^2+|S_8|^2\rr
(\Lambda_{m_1,2 n_1} \Lambda_{m_2,2 n_2}+\Lambda
_{m_1+1/2,2n_1} \Lambda_{m_2+1/2,2 n_2} ) \nonumber\\
&+&\rl |O_8|^2+|C_8|^2
\rr(\Lambda_{m_1,2n_1+1} \Lambda_{m_2,2 n_2+1} + 
\Lambda_{m_1+1/2,2 n_1+1} \Lambda_{m_2+1/2,2n_2+1}) \nonumber\\
&-&\rl V_8\bar{S}_8+S_8\bar{V}_8 \rr(\Lambda_{m_1,2n_1}\Lambda_{m_2+1/2,2n_2}+
\Lambda_{m_1+1/2,2n_1} \Lambda_{m_2,2n_2}) \nonumber\\
&-&\rl
O_8\bar{C}_8+C_8\bar{O}_8\rr(\Lambda_{m_1,2n_1+1}
\Lambda_{m_2+1/2,2n_2+1}+\Lambda_{m_1+1/2,2n_1+1}\Lambda_{m_2,2n_2+1}) \
. \label{e7}
\ea
 
Next we consider a different orientifold projection $\Omega ' = \Omega
(-1)^{f_L} \Pi_1 \Pi_2$ containing parities in the two internal
coordinates. The Klein bottle amplitude

\ba
\mathcal{K}&=&\frac{1}{2} \ \biggl[ (V_8-S_8)
W_{2n_1}W_{2n_2}-(O_8-C_8) W_{2n_1+1} W_{2n_2+1} \biggr] \  \label{e8}
\ea
introduces ${O7}_{+}$-planes localized at $(0,0)$ and ${\overline{O7}}_{-}$
-planes in $(\pi R, \pi R)$ in the $(X_1,X_2)$ plane, which is the type
of system we were searching for. 
The tachyonic scalar is again eliminated by the projection. 
The lowest possible mass scalar states have a degeneracy of four  
and their mass 
\ba
&& M_{T_i}^2 = ({1 \over R_1}- {R_1 \over \alpha'})^2 +({R_2 \over
  \alpha'})^2 \  , \nonumber \\
&& M_{S_i}^2 = ({1 \over R_2}- {R_2 \over \alpha'})^2 +({R_1 \over
\alpha'})^2 \ , 
\label{e9}
\ea
where $i = 1 \cdots 4$, is positive for any values of the compact
radii. 

To cancel the R-R
tadpoles we need therefore to introduce N=32 D7-branes.
The open string amplitudes for branes on top of the $O7_{+}$ planes at
the origin are
\ba
\mathcal{A} &=& \frac{N^2}{2} \ (V_8-S_8)
(W_{2n_1}W_{2n_2}+W_{2n_1+1}W_{2n_2+1})
\ , \nonumber \\
\mathcal{M} &=& -\frac{N}{2} \
\biggl[ \hat{V}_8 (W_{2n_1}W_{2n_2}+ W_{2n_1+1}W_{2n_2+1})- 
\hat{S}_8 (W_{2n_1}W_{2n_2}+W_{2n_1+1}W_{2n_2+1}) \biggr] \ . \label{e10}
\ea
For this particular configuration the gauge group is $SO(32)$, whereas
putting the branes on top of the ${\overline{O7}}_{-}$ planes would
produce an $USp(32)$ gauge group. The second case is the one relevant
for the cosmological solutions discussed in \cite{dmt1,dmt2} and in
Section 5 of the present paper. The absence of the antisymmetric
tensor in this case is easily explained by the fact that the geometric
interpretation needs no T-dualities, and therefore the geometry is
represented by the original twisted torus (\ref{e2}).

\subsection{A chiral model in six dimensions}

The string vacua considered so far, if dimensionally reduced to lower
dimensions, contain non-chiral fermions. 
Even if we are clearly interested in chiral four dimensional
vacua, six-dimensional ones are the next step in this
direction.

 The string vacua we consider in the following  contain D9 and D5 branes,
 and O9 and O5 planes which are actually $O5_{+}$-${\overline{O5}}_{-}$
 systems. By taking three T-dualities along directions transverse to
 the D5 branes, we find configurations consisting of  D8 branes
 and $O8_{+}$-${\overline{O8}}_{-}$ systems, together with the BPS D6 and $O6_{+}$ 
 configurations. If the D6 branes are placed democratically on top of the $O6_{+}$
 planes, the resulting system has no couplings to the massless
 closed string modes and the effective lagrangian and the
 corresponding classical solution are precisely the one worked out in
 \cite{dmt1}.  

The construction of the model starts with a Scherk-Schwarz deformation
of the SUSY model $T^4/\mathbb{Z}_2$ by $(-1)^{F}\times \delta$,
with $F$ the spacetime fermion number and $\delta$ the shift $\delta X_9=X_9+\pi R_9$.
The $\mathbb{Z}_2$ acts by convention in the $(X_6,X_7,X_8,X_9)$ coordinates.
The torus amplitude of the model is:

\ba \mathcal{T}&=& \frac{1}{4}\pl 
|V_8-S_8|^2\Lambda_{m,n}+|V_8+S_8|^2(-1)^m\Lambda_{m,n}\pr
\Lambda^{(3,3)}+\nonumber \\
&+&\frac{1}{4}\pl |O_8-C_8|^2\Lambda_{m,n+1/2}+|O_8+C_8|^2(-1)^m\Lambda_{m,n+1/2}\pr
\Lambda^{(3,3)} + \nonumber \\
&+&\frac{1}{4}\pl\rl |Q_o-Q_v|^2+|Q'_o-Q'_v|^2\rr \al
\frac{2\eta}{\theta_2}\ar ^4+16 \rl \al Q_s+Q_c \ar ^2+\al Q'_s+Q'_c \ar ^2\rr
\al \frac{\eta}{\theta_4}\ar^4\pr +\nonumber\\
&+&\frac{1}{4}\pl16\rl \al Q_s-Q_c \ar ^2+\al Q'_s-Q'_c \ar ^2\rr
\al \frac{\eta}{\theta_3}\ar^4\pr \ , \label{c1}
\ea
where:
\ba
&&Q_o=V_4O_4-C_4C_4 \quad \quad , \quad \quad
Q'_o=V_4O_4-S_4S_4 \ , \nonumber \\
&&Q_v=O_4V_4-S_4S_4\quad \quad , \quad \quad Q'_v=O_4V_4-C_4C_4 \ , 
\nonumber \\
&&Q_s=O_4C_4-S_4O_4\quad \quad , \quad \quad Q'_s=O_4S_4-C_4O_4 \ , \nonumber\\
&&Q_c=V_4S_4-C_4V_4\quad \quad , \quad \quad Q'_c=V_4C_4-S_4V_4 \ . \label{c2}
\ea
The three lattice sums $\Lambda^{(3,3)}$ refer to the $(X_6,X_7,X_8)$
coordinates, whereas the remaining lattice sums ($\Lambda_{m,2n}$, etc)
refer to the coordinate $X_9$ involved in the supersymmetry breaking deformation. 
After the rescaling $R_9 \rightarrow 2 R_9 $ the torus amplitude becomes:

\ba 
\mathcal{T}&=& \frac{1}{2}\pl \rl |V_8|^2+|S_8|^2\rr
\Lambda_{m,2n}+\rl |O_8|^2+|C_8|^2 \rr
\Lambda_{m,2n+1}\pr\Lambda^{(3,3)}-\nonumber\\ 
&-& \frac{1}{2}\pl \rl V_8\bar{S}_8+S_8\bar{V}_8\rr
\Lambda_{m+1/2,2n} + \rl O_8\bar{C}_8+ C_8\bar{O}_8\rr
\Lambda_{m+1/2,2n+1}\pr \Lambda^{(3,3)}+\nonumber\\ 
&+& \frac{1}{4}\pl \rl |Q_o-Q_v|^2+|Q'_o-Q'_v|^2\rr
 \al \frac{2\eta}{\theta_2}\ar ^4+16 \rl \al Q_s+Q_c \ar ^2+\al
 Q'_s+Q'_c \ar ^2\rr 
\al \frac{\eta}{\theta_4}\ar^4\pr +\nonumber\\
&+&\frac{1}{4}\pl16\rl \al Q_s-Q_c \ar ^2+\al Q'_s-Q'_c \ar ^2\rr
\al \frac{\eta}{\theta_3}\ar^4\pr \ . \label{c3}
\ea

 Next we construct the orientifold by gauging the discrete symmetry
 $\Omega'=\Omega (-1)^{f_L} $, where $\Omega$ is the standard
 worldsheet parity operator and $(-1)^{f_L} $ is the worldsheet
 fermion number\footnote{The Scherk-Schwarz orientifold builded
 with the standard $\Omega$ projection was constructed in
 \cite{ads1}.}. The corresponding Klein bottle amplitude is then:

\ba 
\mathcal{K}&=&\frac{1}{4}\pl
(V_8-S_8)(P_mP^3+W_{2n}W^3)-(O_8-C_8)W_{2n+1}W^3\pr +\nonumber\\&+& 
\frac{2\times 8}{4}(Q_s+Q_c-Q'_s-Q'_c)\rl\frac{\eta}{\theta_4}\rr^4 \ . \label{c4} 
\ea
Notice that the negative parity of the untwisted sector tachyon in
(\ref{c4}) is accompanied by a peculiar orientifold action in the
twisted sector : the twisted sector is symmetrized in half of the
fixed points, while it is antisymmetrized in the other half.  
 
The untwisted closed string spectrum of the model has only massive
fermions due to the Scherk-Schwarz deformation. The massless 
spectrum contains the bosons from the gravity multiplet, one tensor
multiplet and four hyper-multiplets. 
The sixteen fixed points of the twisted sector contain supersymmetric
multiplets. There are eight twisted hypers localized in eight fixed
points and eight twisted tensor multiplets in the remaining eight fixed points.
 
To determine the content in O5-planes of the model we look
at the transverse Klein bottle amplitude:

\ba \widetilde{\mathcal{K}}&=& \frac{2^5}{4}\pl \rl v W_eW_e^3+\frac{1}{2v}PP_e^3\rr \rl
V_8-S_8 \rr -\frac{1}{2v}(-1)^mP_mP_e^3(V_8+S_8)\pr\nonumber \\
&+&\frac{2^5}{4}\pl \rl Q_o-Q_v-Q'_o+Q'_v \rr
\rl\frac{2\eta}{\theta_2}\rr^2\pr \ . \label{c5}
\ea

 The model contains $O9_{+}$ planes, 16 $O5_{+}$ and 16
 $\overline{O5}_{-}$ planes. The RR tadpole cancellation requires
 32 $D9$ branes and 32 $D5$ branes. 

 Let us start with the case where all D5 branes are
 coincident with one $O5_{+}$ plane. The transverse annulus amplitude
 reads :

 \ba \widetilde{\mathcal{A}}&=& \frac{2^{-5}}{4} v\pl 
 (N+\bar{N})^2(V_8-S_8)W_{2n}-(N-\bar{N})^2(O_8-C_8)W_{2n+1}\pr W^3+
 \nonumber\\ 
&+& \frac{2^{-5}}{4}
\pl8(R_N+R_{\bar{N}})^2(Q_s+Q_c)\rl\frac{\eta}{\theta_4}\rr^2 -
8(R_N-R_{\bar{N}})^2(Q'_s+Q'_c)\rl\frac{\eta}{\theta_4}\rr^2
\pr+\nonumber \\
&+&  \frac{2^{-5}}{4} \pl
\frac{D^2}{v}P^4(V_8-S_8)+2(N+\bar{N})D(Q_o-Q_v)\rl\frac{2\eta}{\theta_2}\rr^2
+16R_D^2(Q_s+Q_c)\rl\frac{\eta}{\theta_4}\rr^2\pr+\nonumber\\
&+& \frac{2^{-5}}{4} \pl
2(R_N+R_{\bar{N}})R_D(Q_s-Q_c)\rl\frac{2\eta}{\theta_3}\rr^2 \pr, \label{c6} 
\ea
where $N+\bar{N}$ is the number of $D_9$ branes, $D$ is the number
of the  $D_5$ branes, while $R_N, R_{\bar{N}}, R_D$ encode the
orbifold action of $\mathbb{Z}_2$ on the Chan Paton charges. 

The transverse M{\"o}bius amplitude is obtained from factorization of 
$\widetilde{K}$ and $\widetilde{A}$ : 

\ba 
\widetilde{\mathcal{M}}&=&\frac{1}{2}\pl
-(N+\bar{N})v[\hat{V}_8(-1)^n W_{2n}-\hat{S}_8 W_{2n}] W_e^3+
\frac{D}{v} \hat{S}_8 P_e P_e^3 \pr
+ \nonumber\\
&+&\frac{1}{2}\pl -
D(\hat{Q}_o-\hat{Q}_v)\rl\frac{2\hat{\eta}}{\hat{\theta}_2}\rr^2-\frac{D}{v}\hat{V}_8
P_{2m+1} P_e^3-(N+\bar{N})
(\hat{S}_4 \hat{S}_4-\hat{C}_4
\hat{C}_4)\rl\frac{2\hat{\eta}}{\hat{\theta}_2}\rr^2\pr \ , \label{c7}
\ea

and the R-R tadpoles condition imply  $N+\bar{N}=D=32$ and $R_N=R_{\bar{N}}=R_D=0$.  

 The direct amplitudes for open strings are obtained from the
 transverse channel by an $S$ transformation for the annulus and a $P$ 
transformation for the M{\"o}bius amplitude:

\ba 
\mathcal{A} &=& \frac{1}{8}\pl(
N+\bar{N})^2(V_8-S_8)(P_m+P_{m+1/2})-(N-\bar{N})^2(V_8+S_8)(P_m-P_{m+1/2})\pr
P^3 + \nonumber\\
&+& \frac{1}{8}\pl
2D^2(V_8-S_8)W^4+4(N+\bar{N})D(Q_s+Q_c)\rl\frac{\eta}{\theta_4}\rr^2\pr
+\nonumber\\
&+& \frac{1}{8}\pl (R_N+R_{\bar{N}})^2(Q_o-Q_v)\rl\frac{2\eta}{\theta_2}\rr^2-
     (R_N-R_{\bar{N}})^2 (Q'_o-Q'_v)\rl\frac{2\eta}{\theta_2}\rr^2  \pr
     +\nonumber\\
&+&  \frac{1}{8}\pl 2R_D^2(Q_o-Q_v)\rl\frac{2\eta}{\theta_2}\rr^2
+4(R_N+R_{\bar{N}})R_D(Q_s-Q_c)\rl\frac{\eta}{\theta_3}\rr^2 \pr \ ,
\nonumber \\
\mathcal{M}&=& \frac{1}{4}\pl
-(N+\bar{N})(\hat{V}_8 P_{m+1/2}-\hat{S}_8 P_m)
P^3-D(\hat{V}_8(-1)^n-\hat{S}_8) W_nW^3\pr+\nonumber\\
&+&
\frac{1}{4}\pl(N+\bar{N})(\hat{S}_4\hat{S}_4-\hat{C}_4\hat{C}_4)\rl
\frac{2\hat{\eta}}{\hat{\theta}_2}\rr^2+
D (\hat{Q}_o-\hat{Q}_v) \rl
\frac{2\hat{\eta}}{\hat{\theta}_2}\rr^2\pr \ . \label{c8}
\ea

 In order to have a consistent particle interpretation of the open
 amplitudes we have to parameterize $N,\bar{N},D, R_N,
 R_{\bar{N}},R_D$ in terms of the real Chan Paton multiplicites:
\ba 
N &=& n_1+n_2\quad  \quad, \quad \quad
\quad \quad \quad D=d+\bar{d} \ , \nonumber\\
R_N&=&i(n_1-n_2)\quad , \quad \quad  \quad \quad R_D=i(d-\bar{d}) \
. \label{c9}
\ea

The gauge group is $[U(8)\otimes U(8)]_9 \otimes U(16)_5$. The D9
massless spectrum is nonsupersymmetric and consists of 4 scalars in the
$({\bf 8,\bar{8}, 1})+({\bf \bar{8},8,1})$. There are fermions (gaugini) in the
bifundamental ({\bf 8,8,1}) and $({\bf \bar{8},\bar{8},1})$
representations. Matter fermions (of opposite chirality) are in
$({\bf 28,1,1})+({\bf \overline{28},1,1})+({\bf 1,28,1})+({\bf
1,\overline{28},1})$. The D5 massless spectrum is supersymmetric and
has, in addition to the adjoint vector multiplet, hypermultiplets in 
$({\bf 1,1,120})+({\bf 1,1,\overline{120}})+
({\bf 8,1,\overline{16}})+({\bf 1,\bar{8},\overline{16}})$. 
The model we just constructed has all D5 branes on top of one $O5_{+}$
-plane. The massless spectrum on D5 branes should therefore be
supersymmetric and this is indeed the case.  
  Notice from the Chan-Paton parameterization (\ref{c9}) the consistency
 of the string amplitudes (\ref{c6})-(\ref{c7}) in the closed channel. 
Indeed, it is well known from field theory arguments that in six
 dimensions branes cannot couple to twisted hypermultiplets. 
 In (\ref{c7}) these couplings are proportional to $R_N + R_{\bar N}$ and
 $R_D$, and are indeed unphysical. On the other hand, branes can
 consistently couple to (the Hodge dual of) twisted tensor multiplets,
 and indeed in (\ref{c7}) the couplings of D9 branes (D5 branes in this
 model live in one fixed point containing one hyper), proportional to
 $R_N - R_{\bar N}$, are physical. 
    
Irreducible gauge and gravitational anomalies are easily
seen to cancel in this model. The reducible part of the anomaly polynomial is
\be
I_8 = - {1 \over 16} ({\rm tr} R^2-{\rm tr} F_1^2-{\rm tr} F_2^2-
{\rm tr} F_5^2 )^2 
- {1 \over 4} ({\rm tr} F_1^2-{\rm tr} F_2^2)^2
+ {1 \over 16} ({\rm tr} F_1^2+{\rm tr} F_2^2- {\rm tr} F_5^2 )^2 \ , \label{c11}
\ee
and is taken care by the generalized Green-Schwarz mechanism \cite{aug}.

The geometric configurations for which the time-dependent solutions are valid
correspond to moving some D5 branes on top of $O5_{+}$ planes and some
on top of ${\overline{O5}}_{-}$ planes. Moving D5 branes from
a fixed point containing closed twisted hypers to another one
containing twisted tensors presents interesting subtleties in
that the cylinder amplitude, even if a Wilson line deformation of the
former one, leads to an amplitude that has a qualitatively different 
structure. For example,
moving all D5 branes on top of a  ${\overline{O5}}_{-}$ plane gives a
tree-level cylinder amplitude:
 \ba \widetilde{\mathcal{A}}&=& \frac{2^{-5}}{4} v\pl 
 (N+\bar{N})^2(V_8-S_8)W_{2n}-(N-\bar{N})^2(O_8-C_8)W_{2n+1}\pr W^3+
 \nonumber\\ 
&+& \frac{2^{-5}}{4}
\pl8(R_N+R_{\bar{N}})^2(Q'_s+Q'_c)\rl\frac{\eta}{\theta_4}\rr^2 -
8(R_N-R_{\bar{N}})^2(Q_s+Q_c)\rl\frac{\eta}{\theta_4}\rr^2
\pr+\nonumber \\
&+&  \frac{2^{-5}}{4} \pl
\frac{D^2}{v}P^4(V_8-S_8)+2(N+\bar{N})D(Q_o-Q_v)\rl\frac{2\eta}{\theta_2}\rr^2
+16R_D^2(Q'_s+Q'_c)\rl\frac{\eta}{\theta_4}\rr^2\pr+\nonumber\\
&+& \frac{2^{-5}}{4} \pl
2(R_N+R_{\bar{N}}) R_D
(-O_4S_4-C_4O_4+V_4C_4+S_4V_4)\rl\frac{2\eta}{\theta_3}\rr^2 \pr \ ,  \label{c12} 
\ea
which roughly speaking exchanges twisted sector characters $Q_{s,c} 
\leftrightarrow Q'_{s,c}$ 
and generates a completely nonsupersymmetric character in the last
line. The loop open amplitudes in this case are given by:

\ba 
\mathcal{A} &=& \frac{1}{8}\pl(
N+\bar{N})^2(V_8-S_8)(P_m+P_{m+1/2})-(N-\bar{N})^2(V_8+S_8)(P_m-P_{m+1/2})\pr
P^3 + \nonumber\\
&+& \frac{1}{8}\pl
2D^2(V_8-S_8)W^4+4(N+\bar{N})D(Q_s+Q_c)\rl\frac{\eta}{\theta_4}\rr^2\pr
+\nonumber\\
&+& \frac{1}{8}\pl (R_N+R_{\bar{N}})^2(Q'_o-Q'_v)\rl\frac{2\eta}{\theta_2}\rr^2-
     (R_N-R_{\bar{N}}^2)(Q_o-Q_v)\rl\frac{2\eta}{\theta_2}\rr^2  \pr
     +\nonumber\\
&+&  \frac{1}{8}\pl 2R_D^2(Q'_o-Q'_v)\rl\frac{2\eta}{\theta_2}\rr^2
+4(R_N+R_{\bar{N}})R_D (-O_4C_4+V_4S_4-S_4O_4+C_4V_4)\rl\frac{\eta}{\theta_3}\rr^2 \pr \ ,
\nonumber \\
\mathcal{M}&=& \frac{1}{4}\pl
-(N+\bar{N})(\hat{V}_8 P_{m+1/2}-\hat{S}_8 P_m)
P^3+ D \ (\hat{V}_8(-1)^n + \hat{S}_8) W_nW^3\pr+\nonumber\\
&+&
\frac{1}{4}\pl(N+\bar{N})(\hat{S}_4\hat{S}_4-\hat{C}_4\hat{C}_4)\rl
\frac{2\hat{\eta}}{\hat{\theta}_2}\rr^2+ D (\hat{Q}_o-\hat{Q}_v) \rl
\frac{2\hat{\eta}}{\hat{\theta}_2}\rr^2\pr \ , \label{c13}
\ea
and the Chan Paton parameterization and RR tadpole conditions in this case are
\ba 
N &=& n_1+n_2 = 32 \quad , \quad \quad \quad \quad D=d_1+ d_2 = 32 \ , \nonumber\\
R_N&=& n_1- n_2 = 0 \quad , \quad \quad \quad \quad R_D= d_1- d_2  = 0 \
. \label{c14}
\ea
Notice again that the couplings to the twisted tensors $Q'_{s,c}$ are
physical, whereas the ones to the twisted hypers  $Q_{s,c}$ are
unphysical.  The amplitudes (\ref{c13}) describe a configuration with
all D5 branes on top of an ${\overline{O5}}_{-}$ plane and the
gauge group becomes $[U(8)\otimes U(8)]_9 \otimes [USp(16)
\otimes USp(16)]_5$. As expected, in
this case both D9 and D5 massless spectra are nonsupersymmetric. 
The massless spectrum consists of 4 scalars in the
$({\bf 8,\bar{8}, 1,1})+({\bf \bar{8},8,1,1})+ ({\bf 1,1,16,16})+ ({\bf 8,1,1, 16})+ ({\bf
1,8,16, 1})$. There are fermions (gaugini) in the
bifundamental $({\bf 8,8,1,1})+ ({\bf \bar{8},\bar{8},1,1})+({\bf 1,1,16,16})$ 
representations. Matter fermions (of opposite chirality) are in
$({\bf 28,1,1,1})+({\bf \overline{28},1,1,1})+ ({\bf 1,28,1,1})+({\bf
1,\overline{28},1,1})+({\bf 1,1,120,1})+({\bf 1,1,1,120})+
({\bf 8,1,16,1})+({\bf 1,8,1, 16}) $. There are 2 goldstinos on D5
branes, in agreement with the considerations of \cite{dm2}. 
By denoting $G_{1,2}$ the $USp(16)_i$ D5 gauge factors, the anomaly
polynomial for this string vacuum is
\ba
&& I_8 = - {1 \over 16} ({\rm tr} R^2-{\rm tr} F_1^2-{\rm tr} F_2^2-
{1 \over 2} {\rm tr} G_1^2- {1 \over 2} {\rm tr} G_2^2 )^2 
+ {1 \over 16}  ({\rm tr} F_1^2+{\rm tr} F_2^2-
{1 \over 2} {\rm tr} G_1^2- {1 \over 2} {\rm tr} G_2^2 )^2 \nonumber \\
&&-{1 \over 32}  ({\rm tr} F_1^2-{\rm tr} F_2^2+
2 {\rm tr} G_1^2- 2 {\rm tr} G_2^2 )^2 - {7 \over 32}  ({\rm tr} F_1^2-{\rm tr} F_2^2)^2 
 \  \label{c15}
\ea
and is taken care again by the generalized Green-Schwarz mechanism \cite{aug}.
 
Moving only part of the D5 branes from the origin to the twisted tensor
fixed points produce a D5 branes gauge group $[U(n)
\otimes USp(16-n) \otimes USp(16-n)]_5$.

\section{Four-dimensional $Z_2 \times Z_2$ vacua}

We now turn to four dimensional compactification on \mbox{$T^{2}\times
T^{2}\times T^{2}$} of the  Type-IIB theory, orbifolded by the
\mbox{$Z_{2}\times Z_{2}$} action generated by the identity (we will
call it ``$o$'') and the $\pi$ rotations $g:(+,-,-)$, $f:(-,+,-)$,
$h:(-,-,+)$, where the three entries in the parentheses refer to the
three internal tori, while ``$+$'' and ``$-$'' denote the two group elements of $Z_{2}$.
We deform the resulting \mbox{$Z_{2}\times Z_{2}$} model 
by $(-1)^{F}\times \delta$, with $F$ the spacetime fermion number and
$\delta$ the shift $\delta \ X_9 = X_9+\pi R_9$.

The torus partition function, after a standard rescaling of the $X_9$
coordinate in order to go to the Scherk-Schwarz basis, is given by :
\ba 
\mathcal{T} &=& \frac{1}{4} \biggl\{ \Lambda_{1} \Lambda_{2} \Lambda_{m,n} \left[ \Lambda_{m,2n}
(|T^{B}_{oo}|^{2}+|T^{F}_{oo}|^{2}) +\Lambda_{m,2n+1}
(|S^{B}_{oo}|^{2}+|S^{F}_{oo}|^{2}) +
\Lambda_{m+1/2,2n}(T^{B}_{oo}\bar{T}^{F}_{oo} +
\bar{T}^{B}_{oo}T^{F}_{oo}) \right.  \nonumber \\
&&\left. + \Lambda_{m+1/2,2n+1} (S^{B}_{oo}\bar{S}^{F}_{oo}+\bar{S}^{B}_{oo}S^{F}_{oo}) \right]
+ \Lambda_{1}| \frac{4\eta^{2}}{\theta^{2}_{2}}|^{2}\frac{1}{2}\left[
  |T_{og}|^{2}+|T^{'}_{og}|^{2}\right] +
\Lambda_{2} |\frac{4\eta^{2}}{\theta^{2}_{2}}|^{2}\frac{1}{2} \left[|T_{of}|^{2}+
|T^{'}_{of}|^{2} \right]   \nonumber\\
&& + |\frac{4\eta^{2}}{\theta^{2}_{2}}|^{2} \Lambda_{m,n} \left[ \Lambda_{m,2n}
(|T^{B}_{oh}|^{2}+|T^{F}_{oh}|^{2})+
\Lambda_{m,2n+1}(|S^{B}_{oh}|^{2}+|S^{F}_{oh}|^{2})+ \Lambda_{m+1/2,2n}
(T^{B}_{oh}\bar{T}^{F}_{oh}+ \bar{T}^{B}_{oh}T^{F}_{oh}) \right.  \nonumber\\
&& \left. +\Lambda_{m+1/2,2n+1} (S^{B}_{oh}\bar{S}^{F}_{oh}+\bar{S}^{B}_{oh}S^{F}_{oh})\right] +
\Lambda_{1}|\frac{4\eta^{2}}{\theta^{2}_{4}}|^{2}\frac{1}{2}\left[|T_{go}|^{2}+|T^{'}_{go}|^{2}\right] + \Lambda_{2}|\frac{4\eta^{2}}{\theta^{2}_{4}}|^{2}\frac{1}{2}\left[|T_{fo}|^{2}+|T^{'}_{fo}|^{2}\right] \nonumber\\
&& + |\frac{4\eta^{2}}{\theta^{2}_{4}}|^{2}\Lambda_{m,n} \left[
\Lambda_{m,2n}(|T^{B}_{ho}|^{2}+|T^{F}_{ho}|^{2})+
\Lambda_{m,2n+1}(|S^{B}_{ho}|^{2}+|S^{F}_{ho}|^{2})+\Lambda_{m+1/2,2n}
(T^{B}_{ho}\bar{T}^{F}_{ho}+ \bar{T}^{B}_{ho}T^{F}_{ho}) \right.  \nonumber\\
&&\left. +\Lambda_{m+1/2,2n+1} (S^{B}_{ho}\bar{S}^{F}_{ho}+\bar{S}^{B}_{ho}S^{F}_{ho})\right] +
\Lambda_{1}|\frac{4\eta^{2}}{\theta^{2}_{3}}|^{2}\frac{1}{2}\left[|T_{gg}|^{2}+|T^{'}_{gg}|^{2}\right] + \Lambda_{2}|\frac{4\eta^{2}}{\theta^{2}_{3}}|^{2}\frac{1}{2}\left[|T_{ff}|^{2}+|T^{'}_{ff}|^{2}\right]  \nonumber\\
&& + |\frac{4\eta^{2}}{\theta^{2}_{3}}|^{2}\Lambda_{m,n} \left[\Lambda_{m,2n}
(|T^{B}_{hh}|^{2}+|T^{F}_{hh}|^{2})+
\Lambda_{m,2n+1}(|S^{B}_{hh}|^{2}+|S^{F}_{hh}|^{2})+\Lambda_{m+1/2,2n}
(T^{B}_{hh}\bar{T}^{F}_{hh}+ \bar{T}^{B}_{hh}T^{F}_{hh}) \right.  \nonumber\\
&& \left. -\Lambda_{m+1/2,2n+1} (S^{B}_{hh}\bar{S}^{F}_{hh}+\bar{S}^{B}_{hh}S^{F}_{hh})\right] +
\epsilon \ 
|\frac{8\eta^{3}}{\theta_{2}\theta_{3}\theta_{4}}|^{2}\frac{1}{2}\left[|T_{gh}|^{2}+|T^{'}_{gh}|^{2}+|T_{gf}|^{2}+|T^{'}_{gf}|^{2}+|T_{hg}|^{2}
\right. \nonumber\\
&&
\left. +|T^{'}_{hg}|^{2}+|T_{hf}|^{2}+|T^{'}_{hf}|^{2}+|T_{fg}|^{2}+|T^{'}_{fg}|^{2}+|T_{fh}|^{2}+|T^{'}_{fh}|^{2}
\right]  \biggr\}, \label{f1}
\ea
where we left implicit the contribution of the transverse bosons and
the argument of the characters, $q=exp(2i\pi\tau)$, with $\tau$ the
modulus of the torus and $\epsilon = \pm 1$. The choice  $\epsilon =
1$  defines the model without discrete torsion, while the choice  $\epsilon =
-1$ defines the model with discrete torsion. With the characters used in
(\ref{f1}), the would-be tachyon is contained in $S_{oo}^B$ and
$S_{oh}^B$ (see their explicit definition in the appendix).  
Moreover, $\Lambda_1$ ($\Lambda_2$) denote the lattice summations in the
first (second) torus, whereas the lattice summations written
separately for the two torus coordinates 
($\Lambda_{m,n} \Lambda_{m,2n} $ ,etc) refer to the third torus, the
first (second) sum referring to the $X_8$ ($X_9$) coordinate.
\subsection{A model without discrete torsion}

We start our four-dimensional constructions by discussing the model
without discrete torsion  $\epsilon = 1$. There are several choices
actually, depending on three signs $\epsilon_i = \pm 1$, where
$\epsilon_i=1$ signals typically the existence of $O5_{+}$ planes,
whereas $\epsilon_i=-1$ that of  $O5_{-}$ planes. The different
possibilities are restricted by the condition
\be
\epsilon \quad = \quad \epsilon_1 \ \epsilon_2 \ \epsilon_3 \ . \label{f01}
\ee
The model relevant  for the cosmological solution discussed in \cite{dmt2}
and in the next section, constructed in this paragraph, has
$\epsilon_i$=1.  In order to consistently
remove the tachyon, we use the orientifold projection $\Omega'=\Omega
(-1)^{f_L}$\footnote{The corresponding orientifold model with the standard projection $\Omega$
and with $\epsilon_i = 1$ was already constructed by A.Cotrone \cite{cotrone}.}.
 
The direct channel Klein bottle amplitude reads:
\ba 
{\mathcal K} &=& \frac{1}{8} \biggl\{ (P_{1}P_{2}P_{3} + (P_{1}W_{2}+W_{1}P_{2}) W_{n} W_{2n} + 
W_{1}W_{2}P_3) T_{oo} - (P_1W_2+ W_{1}P_{2}) W_{n} W_{2n+1} S_{oo}
\nonumber\\ 
&&  + 16\left(\frac{\eta}{\theta_{4}}\right)^{2} \left[ (P_{1}+W_{1})(T_{go}-T^{'}_{go})
+ (P_{2}+W_{2})(T_{fo}- T^{'}_{fo})  \right. \nonumber\\
&& \left. + 2P_{3}T_{ho} +
2 W_{n}(W_{2n}T_{ho}- W_{2n+1} S_{ho}) \right] \biggr\} , \label{f2}
\ea
where $P_{i}$ ($W_{i}$) denotes the restriction of $\Lambda_{i}$ to
its momentum (winding) sublattice.
Notice the peculiar orientifold projection in the twisted sector,
which is symmetrized in half of the orbifold fixed point and
antisymmetrized in the other half, in analogy with the
six-dimensional orbifold constructed in the previous section.

The untwisted massless closed string spectrum consists of the bosonic
part of the $Z_2 \times Z_2$ orientifold one (without discrete
torsion). The twisted spectrum is ${\cal N}=1$ supersymmetric and consists of $32$
chiral multiplets and $16$ vector multiplets distributed in the $48$
fixed points of the orbifold.  The would-be tachyon in the spectrum is again removed by the
orientifold projection, as promised, whereas the lowest mass scalars
have a positive definite mass given by the formula (\ref{s5}).   
  
A modular S-transformation gives the transverse channel Klein bottle amplitude:
\ba 
\mathcal{\widetilde{K}} &=& \frac{2^{5}}{8} \biggl\{ (v_{1}v_{2}W^{e}_{1}W^{e}_{2}+
\frac{1}{v_{1}v_{2}}P^{e}_{1}P^{e}_{2}) v_3W^{e}_{3} T_{oo} \nonumber
\\
&&+ (\frac{v_{1}}{v_{2}v_{3}}W^{e}_{1}P^{e}_{2} + \frac{v_{2}}{v_{1}v_{3}}P^{e}_{1}W^{e}_{2})
P_{2m}(P_{2m+1}T^{B}_{oo} +P_{2m}T^{F}_{oo}) \nonumber\\
&& +\left(\frac{2\eta}{\theta_{2}}\right)^{2} \left[
(v_{1}W^{e}_{1}+\frac{P^{e}_{1}}{v_{1}})(T_{og}-T^{'}_{og}) 
+  (v_{2}W^{e}_{2}+\frac{P^{e}_{2}}{v_{2}})(T_{of}-T^{'}_{of}) +
2v_{3}W^{e}_{3}T_{oh}  \right. \nonumber\\
&& \left. +
2\frac{P_{2m}}{v_{3}}(P_{2m+1}T^{B}_{oh}+P_{2m}T^{F}_{oh}) \right] \biggr\}, \label{f3}
\ea
where $v_{i}$ denote the volumes of the compactified tori.
The nature and the geometry of the orientifold planes is completely
encoded in (\ref{f3}), which describes $O9_{+}$ and 16 $O5_{3,+}$
planes, parallel to the compact dimension used in the Scherk-Schwarz
mechanism, together with 8 $O5_{1,+}-{\overline{O5}}_{1,-}$ and 8
$O5_{2,+}-{\overline{O5}}_{2,-}$ pairs of orientifold planes, orthogonal to the
same compact dimension. The RR tadpole cancellation conditions ask
for a net number of 32 D9 branes, described by the Chan-Paton index $N$
in the following, as well as a net number of 32 $D5_{i}$ five-branes,
of CP index $D_i$. To make contact with the cosmological
solutions of interest for us, we perform three T-dualities, say in
$X_6,X_7$ and $X_8$ coordinates. The D9 branes become D6 and the $D5_{3}$ 
become $D6_{3}$, both of them parallel to $X_9$,  singled out
in our construction. The $D5_{1}$ branes 
become $D8_{1}$ and the $D5_{2}$ branes become $D4_{2}$, both of them
orthogonal to $X_9$. This configuration does not have the 9d solution
of \cite{dmt1} as an exact solution of the classical equations, but by
smearing over the four coordinates perpendicular to the $D5_{2}$ branes
and parallel to the $D8_{1}$ branes we get the same five dimensional
lagrangian \cite{dmt2}, (\ref{n1}).   
  
If we choose the simplest configuration of D-branes at the origin of
the compact space, then the tree-level (transverse) cylinder amplitude
is:
\ba 
\mathcal{\widetilde{A}} &=& \frac{2^{-5}}{8} \biggl\{ v_{1}v_{2}v_{3}W_{1}W_{2} W_{n} [
(N+{\bar N})^2 W_{2n}T_{oo}-(N-{\bar N})^2 W_{2n+1}S_{oo}]  \nonumber\\
&& + \frac{v_{1}}{v_{2}v_{3}}W_{1}P_{2}P_3 D_1^2 T_{oo} 
 +\frac{v_{2}}{v_{1}v_{3}}P_{1}W_{2}P_3 D_2^2 T_{oo}  \nonumber\\
&& + \frac{v_3}{v_1 v_{2}} P_{1}P_{2} W_{n} 
[ (D_3+{\bar D}_3)^2 W_{2n}T_{oo}-(D_3-{\bar D}_3)^2 W_{2n+1}S_{oo}]  \nonumber\\
&&  +2 \left(\frac{2\eta}{\theta_{2}}\right)^{2} \left[ v_1 W_1 (N+{\bar
N}) D_1 T_{og}+   v_2 W_2 (N+{\bar N}) D_2 T_{of}+ \right. \nonumber \\
&& \left.  v_3 W_n W_{2n} (N + {\bar N}) (D_{3}+ {\bar D}_3) T_{oh} -
v_{3} W_n W_{2n+1} (N - {\bar N}) (D_{3}- {\bar D}_3) S_{oh} \right.
 \nonumber \\ &&+ 
\left. \frac{P_1}{v_1} D_2 (D_3 +
{\bar D}_3) T_{og}+ \frac{P_2}{v_2} D_1 (D_3 + {\bar D}_3) T_{of} 
+ \frac{P_3}{v_3} D_1 D_2 T_{oh} \right] \biggr\}. \label{f4}
\ea

The direct channel cylinder, obtained by an S-transformation, is
\ba 
\mathcal{A} &=& \frac{1}{8} \biggl\{ {P_{1}P_{2} \over 2} \left[ (N+ {\bar
N})^{2} (P_m +P_{m+1/2}) T_{oo}
- (N- {\bar N})^{2} (P_m-P_{m+1/2}) T^{B-F}_{oo} \right] 
\nonumber \\ 
&& + (P_{1}W_{2} W_3 D_1^2+ W_1 P_2 W_3 D_2^2) T_{oo}+ \nonumber \\
&& {W_1W_2 \over 2}
\left[ (D_3+ {\bar D}_3)^{2} (P_m +P_{m+1/2}) T_{oo}
-  (D_3- {\bar D}_3)^{2} (P_m -P_{m+1/2}) T^{B-F}_{oo}
\right] \nonumber\\ && +
2 \left(\frac{\eta}{\theta_{4}}\right)^{2} \left[ P_{1}(N+ {\bar N})
D_1 T_{go} + P_{2}(N+{\bar N}) D_2 T_{fo}  \right. \nonumber \\
&& \left. + {1 \over 2} (N+{\bar N}) (D_3+{\bar
D}_3) (P_{m}+P_{m+1/2}) T_{ho} -  {1 \over 2} (N-{\bar N}) (D_3-{\bar D}_3)
(P_{m}-P_{m+1/2}) T^{B-F}_{ho} \right. \nonumber\\
&& \left. + W_1 D_2 (D_3 + {\bar D}_3) T_{go} + W_{2} D_1 (D_3+{\bar D}_3)
T_{fo} + W_3 D_1 D_2 T_{ho} \right]  \biggr\} \ . \label{f5}
\ea
The tree-level channel M{\"o}bius amplitude is found, as usual, by
factorization :
\ba 
\mathcal{{\widetilde M}} &=& -\frac{1}{4} \biggl\{ [(N+ {\bar N})
v_{1}v_{2} W^{e}_1 W^{e}_2 + \frac{D_3 + {\bar D}_3}{v_{1}v_{2}}
P^{e}_1 P^{e}_{2}] v_3 [(-1)^n \hat{T}^{B}_{oo}+\hat{T}^{F}_{oo}] W_{2n}  \nonumber\\
&& + (\frac{v_{1}}{v_{2}v_{3}} D_1 W^{e}_{1} P^{e}_2
+\frac{v_{2}}{v_{1}v_{3}} D_2 P^{e}_{1} W^{e}_2)
(P_{2m} \hat{T}^{F}_{oo}+ P_{2m+1}\hat{T}^{B}_{oo})  \nonumber\\
&& +\left(\frac{2\hat{\eta}}{\hat{\theta}_{2}}\right)^{2} 
\left[ v_{1}W^{e}_{1}( (N+ {\bar N}) \hat{T}^F_{og}+D_1 \hat{T}_{og})+
{P^e_1 \over v_1} ( (D_3+ {\bar D}_3) \hat{T}^F_{og}+D_2 \hat{T}_{og}) \right.   \nonumber\\
&& \left. + v_2 W^{e}_{2}( (N+ {\bar N}) \hat{T}^F_{of}+D_2 \hat{T}_{of})+ 
{P^e_2 \over v_2} ( (D_3+ {\bar D}_3) \hat{T}^F_{of}+D_1 \hat{T}_{of})
\right. \label{f6} \\
&& \left. + v_3 W^e_3 (N+ {\bar N}+D_3+ {\bar D}_3) [-(-1)^n  \hat{T}^B_{oh}+\hat{T}^F_{oh}]
+ {1 \over v_3}  (D_1+D_2) (P_{2m} \hat{T}^F_{oh}+P_{2m+1} \hat{T}^B_{oh})  
\right] \biggr\}, \nonumber 
\ea
where we have used a proper basis of ``hatted characters''.
In order to obtain the direct Mobius amplitude we have to perform a
P-transformation. The final result is
\ba 
&& \mathcal{M}= -\frac{1}{8} \biggl\{ [(N+{\bar N}) P_{1}P_{2} + (D_3+{\bar D}_3)
W_1 W_2] (P_{m+1/2} \hat{T}^{B}_{oo}+P_m \hat{T}^{F}_{oo}) \nonumber \\
&& + (D_1 P_1 W_2 +D_2 W_1 P_2) [(-1)^n
\hat{T}^{B}_{oo}+\hat{T}^{F}_{oo}] W_3  \nonumber\\
&& - \left(\frac{2\hat{\eta}}{\hat{\theta}_{2}}\right)^{2} 
\left[ P_1 ( (N+ {\bar N}) \hat{T}^F_{og}+D_1 \hat{T}_{og})+
W_1 ( (D_3+ {\bar D}_3) \hat{T}^F_{og}+D_2 \hat{T}_{og}) +
P_2 ( (N+ {\bar N}) \hat{T}^F_{of}+D_2 \hat{T}_{of})  \right. \nonumber \\
&& + W_2 ( (D_3+ {\bar D}_3) \hat{T}^F_{of}+D_1 \hat{T}_{of})
\left. -  (N+ {\bar N}+D_3+ {\bar D}_3) (P_{m+1/2}  \hat{T}^B_{oh}- P_m \hat{T}^F_{oh})
\right. \nonumber \\ 
&& \left. + W_3 (D_1+D_2) (\hat{T}^F_{oh}+(-1)^n \hat{T}^B_{oh})
\right] \biggr\} \ .  \label{f7}
\ea
The appropriate spacetime interpretation fixes the correct Chan-Paton
parameterization to be $N = 2 n \ , \ D_i = 2 d_i$.
The RR tadpole cancellation conditions then read:
\be 
n + {\bar n} = d_3 + {\bar d}_3 = 16 \quad , \quad d_1 = d_2 = 16 \ . \label{f8}
\ee

With the new CP factors, the one-loop open string amplitudes are given by:
\ba
\mathcal{A} &=& (n {\bar n} P_{1}P_{2} +  d_3 {\bar d}_3 W_1 W_2) 
(P_m T^{B}_{oo}+ P_{m+1/2} T^{F}_{oo})
+ {1 \over 2} [(n^2 +{\bar n}^2) P_1 P_2 + (d_3^2+ {\bar d}_3^2) W_1 W_2] \times
\nonumber \\
&& \times (P_{m+1/2} T^{B}_{oo}+ P_m T^{F}_{oo}) + 
({d_1^2 \over 2} P_1 W_2 W_3 + {d_2^2 \over 2} W_1 P_2 W_3 ) T_{oo}
\nonumber \\
&& + \left(\frac{\eta}{\theta_{4}}\right)^{2} \biggl\{ [ P_1 (n+ {\bar n}) d_1+ W_1 d_2(d_3
+ {\bar d}_3)] T_{go} 
 + [ P_2 (n+{\bar n}) d_2+  W_2 d_1 (d_3 + {\bar
d}_3)] T_{fo} + \nonumber \\
&& [ (n {\bar d}_3+ {\bar n} d_3) P_{m} + (n d_3+ {\bar n} {\bar d}_3)
P_{m+1/2}] T^B_{ho} \nonumber\\
&& + [ (n d_3+ {\bar n} {\bar d}_3) P_{m} + (n {\bar d}_3+ {\bar n}
d_3) P_{m+1/2}] T^F_{ho} + 2 W_3 d_1 d_2 T_{ho} \biggr\}  \ , \ \nonumber \\
\mathcal{M} &=& -\frac{1}{4} \biggl\{ [(n+{\bar n}) P_{1}P_{2} + (d_3+{\bar d}_3)
W_1 W_2] (P_{m+1/2} \hat{T}^{B}_{oo}+P_m \hat{T}^{F}_{oo}) \nonumber \\
&& + (d_1 P_1 W_2 +d_2 W_1 P_2) [(-1)^n
\hat{T}^{B}_{oo}+\hat{T}^{F}_{oo}] W_3  \nonumber\\
&& - \left(\frac{2\hat{\eta}}{\hat{\theta}_{2}}\right)^{2} 
\left[ P_1 ( (n+ {\bar n}) \hat{T}^F_{og}+d_1 \hat{T}_{og})+
W_1 ( (d_3+ {\bar d}_3) \hat{T}^F_{og}+d_2 \hat{T}_{og}) +
P_2 ( (n+ {\bar n}) \hat{T}^F_{of}+d_2 \hat{T}_{of}) \right. \nonumber \\
&& \left. + W_2 \ ( (d_3+ {\bar d}_3) \hat{T}^F_{of}+d_1 \hat{T}_{of})
-  (n+ {\bar n}+d_3+ {\bar d}_3) (P_{m+1/2}  \hat{T}^B_{oh}- P_m \hat{T}^F_{oh})
\right. \nonumber \\
&& \left. + W_3 (d_1+d_2) (\hat{T}^F_{oh}+(-1)^n \hat{T}^B_{oh})  \right] \biggr\}
. \label{f9}
\ea
 
The gauge group is $U(8)_{9}\otimes USp(16)_{5_{1}} \otimes USp(16)_{5_{2}}
\otimes U(8)_{5_{3}}$. Since we chose all D-branes to be at the
origin of the compact space, we have coincident $D5_i-O5_{i,+}$
configurations. Since the $D5_1$ and $D5_2$ branes are orthogonal to
the coordinate $X_9$ used in the Scherk-Schwarz mechanism, this must
imply that their tree-level spectrum is supersymmetric.
The charged matter spectrum consists in:
\ba
{\rm 1 \ Weyl \ fermion \ in} \quad : \quad && ({\bf 36+\overline{36},1,1,1}) +  
({\bf 1,1,1,36+\overline{36}}) \ , \nonumber \\
{\rm 3 \ Weyl \ fermions \ in} \quad : \quad &&({\bf 28+\overline{28},1,1,1}) +  
({\bf 1,1,1,28+\overline{28}}) \ , \nonumber \\
{\rm 3 \ complex \ scalars \ in} \quad : \quad &&({\bf 64,1,1,1}) +  
({\bf 1,1,1,64}) \ , \nonumber \\
{\rm 3 \ chiral \ multiplets \ in} \quad : \quad &&({\bf 1,120,1,1}) +  
({\bf 1,1,120,1}) \ , \nonumber \\
{\rm 1 \ Weyl \ fermion \ in} \quad : \quad &&({\bf 8,1,1,8}) +  ({\bf \overline{8},1,1,\overline{8}})  
\ , \nonumber \\
{\rm 1 \ complex \ scalar \ in} \quad : \quad &&({\bf 8,1,1,\overline{8}}) +  
({\bf \overline{8},1,1,8}) \ , \nonumber \\ 
{\rm chiral \ multiplets \ in} \quad : \quad &&({\bf 8 +\overline{8},16,1,1}) +  
({\bf 1,1,16, 8+\overline{8}}) +  ({\bf 8 +\overline{8},1,16,1})+
\nonumber \\ 
&&({\bf 1,16,1, 8+\overline{8}}) + 2 \times ({\bf 1,16,16,1})
\ . \label{f10}
\ea
 
The spectrum is non chiral and therefore the model is free of gauge and gravitational anomalies.
As explained in Section 2 and as in the 6d example of the previous
section, models relevant for the cosmological solutions
constructed in Section 5 must contain  $D5_i-\overline{O5}_{i,-}$
configurations. We need therefore to add appropriate Wilson lines to
the present model. By a careful analysis of the NS-NS and RR charges
of various O-planes, it can be readily realized that this is
consistently done by changing the signs of the nonsupersymmetric
bosonic couplings of the $D5_{1,2}$ branes in the tree-level
(transverse) open string amplitudes
(\ref{f4})-(\ref{f6}). Consequently, the massless $D5_{1,2}$ branes
spectrum become nonsupersymmetric, but the gauge group remains the
same. Since the spectrum is still non chiral, we do not display
it explicitly. 
\subsection{A model with discrete torsion}

The case $\epsilon = -1$ in (\ref{f1}) define models with discrete
torsion. It was known long time ago \cite{bianchi} that the
corresponding models based on the $Z_2 \times Z_2$ and without the
Scherk-Schwarz deformation have no supersymmetric solution. It was
later on realized \cite{aaads} that the reason for it is that these
models contain $O5_{i,-}$ planes for each $\epsilon_i = -1$ and therefore consistency conditions
ask for the introduction of $\overline{D5}_{i}$ antibranes and 
supersymmetry is necessarily broken. The main advantage from a model
building point of view in these models compared to the (simpler) ones
without discrete torsion is that they contain 4d chiral fermions. 
The case of most interest for us, satisfying (\ref{f01}) has
$(\epsilon_1,\epsilon_2,\epsilon_3) = (-1,1,1)$. In this case, we
expect configurations of  $\overline{D5}_{1}$ antibranes and
$O5_{1,-}-\overline{O5}_{1,+}$ planes,  ${D5}_{2}$ branes and
$O5_{2,+}-\overline{O5}_{2,-}$ planes and $D5_{3}$ branes and
$O5_{3,+}$ planes, as we will explicitly check in the following. By
choosing appropriate configurations of $D5_{2}$ ($D5_{1}$) branes
(antibranes) on top of the corresponding orientifold plane systems, we
encounter again the non-BPS configurations discussed in Section 2 and
the corresponding cosmological solution. The case
$(\epsilon_1,\epsilon_2,\epsilon_3) = (-1,-1,-1)$, whereas consistent
as a perturbative orientifold construction, contain   $\overline{D5}_{3}$ branes and
$O5_{3,-}$ planes, which lead to more complicated classical solutions.   
 
We skip here the (rather involved) details of the construction and
consistency checks for the various amplitudes.   
The Klein bottle
\ba 
{\mathcal K} &=& \frac{1}{8} \biggl\{ (P_{1}P_{2}P_{3} + (P_{1}W_{2}+W_{1}P_{2}) W_{n} W_{2n} + 
W_{1}W_{2}P_3) T_{oo} - (P_1W_2+ W_{1}P_{2}) W_{n} W_{2n+1} S_{oo}
\nonumber\\ 
&&  + 16\left(\frac{\eta}{\theta_{4}}\right)^{2} \left[ - (P_{1}-W_{1})(T_{go}-T^{'}_{go})
+ (P_{2}-W_{2})(T_{fo}- T^{'}_{fo})  \right. \nonumber\\
&& \left. + 2P_{3}T_{ho} -
2 W_{n}(W_{2n}T_{ho}- W_{2n+1} S_{ho}) \right] \biggr\} , \label{f11}
\ea
which removes, as in all the other previous models, the closed string tachyon,
contains no massless propagation for twisted fields, as is standard for models
with discrete torsion. 

By introducing appropriate Chan-Paton factors and a diagonal orbifold action
on them, the loop channel cylinder amplitude, in the simplest case
where all the branes are at the origin of the compact space, is given by
\ba 
\mathcal{A} &=& \frac{1}{8} \biggl\{ {P_{1}P_{2} \over 2} \left[ (N_o+ {\bar
N}_o)^{2} (P_m +P_{m+1/2}) T_{oo}
- (N_o- {\bar N}_o)^{2} (P_m-P_{m+1/2}) T^{B-F}_{oo} \right] 
\nonumber \\ 
&& + (P_{1}W_{2} W_3 D_{go}^2+ W_1 P_2 W_3 D_{fo}^2) T_{oo}+ \nonumber \\
&& {W_1W_2 \over 2}
\left[ (D_{ho}+ {\bar D}_{ho})^{2} (P_m +P_{m+1/2}) T_{oo}
-  (D_{ho}- {\bar D}_{ho})^{2} (P_m -P_{m+1/2}) T^{B-F}_{oo}
\right] \nonumber\\ && +
2 \left(\frac{\eta}{\theta_{4}}\right)^{2} \left[ P_{1}(N_o+ {\bar N}_o)
D_{go} T'_{go} + P_{2}(N_o+{\bar N}_o) D_{fo} T_{fo}  \right. \nonumber \\
&& \left. + {N_o+{\bar N}_o \over 2} (D_{ho}+{\bar
D}_{ho}) (P_{m}+P_{m+1/2}) T_{ho} -  {N_o-{\bar N}_o \over 2} (D_{ho}-{\bar D}_{ho})
(P_{m}-P_{m+1/2}) T^{B-F}_{ho} \right. \nonumber\\
&& \left. + W_1 D_{fo} (D_{ho} + {\bar D}_{ho}) T_{go} + W_{2} D_{go} 
(D_{ho} +{\bar D}_{ho}) T'_{fo} + W_3 D_{go} D_{fo} S_{ho} \right]  \nonumber \\
&& - 2 \left(\frac{\eta}{\theta_{3}}\right)^{2} \left[ P_{1}(N_g+ {\bar N}_g)
D_{gg} T^{' B-F}_{gg} + P_{2}(N_f+{\bar N}_f) D_{ff} T_{ff}  \right. \nonumber \\
&& \left. + {N_h+{\bar N}_h \over 2} (D_{hh}+{\bar
D}_{hh}) (P_{m}+P_{m+1/2}) T_{hh} -  {N_h-{\bar N}_h \over 2} (D_{hh}-{\bar D}_{hh})
(P_{m}-P_{m+1/2}) T^{B-F}_{hh} \right. \nonumber\\
&& + \left. W_1 D_{fg} (D_{hg} + {\bar D}_{hg}) T_{gg} + W_{2} D_{gf} 
(D_{hf} +{\bar D}_{hf}) T^{'B-F}_{ff} + W_3 D_{gh} D_{fh} S_{hh} \right] \nonumber \\
&&+ \left.   \left(\frac{2 \eta}{\theta_{2}}\right)^{2} \left[ P_{1}( {1 \over
 2} (N_g+ {\bar N}_g)^2 + D_{gg}^2)  +  W_{1}( {1 \over
 2} (D_{hg}+ {\bar D}_{hg})^2 + D_{fg}^2) \right] T_{og}
\right. \nonumber \\
&&-  \left.  {1 \over 2}  \left(\frac{2\eta}{\theta_{2}}\right)^{2}
  \left[ P_{1} (N_g- {\bar N}_g)^2   +  W_{1} (D_{hg}- {\bar
D}_{hg})^2  \right] T'_{og} \right. \nonumber \\
&&+ \left.  \left(\frac{2\eta}{\theta_{2}}\right)^{2} \left[ P_{2}( {1 \over
 2} (N_f+ {\bar N}_f)^2 + D_{ff}^2)  +  W_{2}( {1 \over
 2} (D_{hf}+ {\bar D}_{hf})^2 + D_{gf}^2) \right] T_{of}
\right. \nonumber \\
&&-  \left.  {1 \over 2}  \left(\frac{2\eta}{\theta_{2}}\right)^{2}
  \left[ P_{2} (N_f- {\bar N}_f)^2   +  W_{2} (D_{hf}- {\bar
D}_{hf})^2  \right] T'_{of} \right. \nonumber \\
&&+ \left.   \left(\frac{2\eta}{\theta_{2}}\right)^{2} \left[{1 \over
 2} (P_{m}+P_{m+1/2}) ((N_h+ {\bar N}_h)^2 + (D_{hh}+{\bar D}_{hh})^2)
+  W_{3} (D_{gh}^2+ D_{fh}^2) \right] T_{oh}
\right. \nonumber \\
&&- \left.   \left(\frac{2\eta}{\theta_{2}}\right)^{2} {1 \over
 2} (P_{m}-P_{m+1/2}) \left[ ((N_h- {\bar N}_h)^2 + (D_{hh}-{\bar D}_{hh})^2)
 \right] T^{B-F}_{oh} \right. \nonumber \\
&& \left. +  {4i \eta^3 \over \theta_2 \theta_3 \theta_4} 
[ (N_g + \overline{N}_g) (-D_{fg} T_{fg} +  
{D_{hg} + \overline{D}_{hg} \over 2} T_{hg}) 
- {1 \over 2}  (N_g - \overline{N}_g)(D_{hg} - \overline{D}_{hg})
T_{hg}^{B-F} \right. \nonumber \\  
&& \left. -  (N_f + \overline{N}_f) (D_{gf} T_{gf}^{'B-F} +  
{D_{hf} + \overline{D}_{hf} \over 2} T_{hf}) +  {1 \over 2}  (N_f -
\overline{N}_f) (D_{hf} - \overline{D}_{hf}) T_{hf}^{B-F} \right. \nonumber \\ 
&& \left. + (N_h + \overline{N}_h) (D_{gh} T_{gh}^{'} -  
D_{fh} T_{fh}) \right. \nonumber \\ 
&& \left. - D_{gg} (D_{hg}+ \overline{{D}_{hg}}) T_{fg}^{'B-F} +
D_{gg} D_{fg} S_{hg} -D_{ff} D_{gf} S_{hf} \right. \nonumber \\
&& \left. -  D_{ff} (D_{hf}+ \overline{{D}_{hf}}) T_{gf} +  
(D_{hh}+ \overline{{D}_{hh}}) (D_{fh} T_{gh}- D_{gh} T_{fh}^{'})
] \biggr\}\right. \label{f12}\ea
The loop channel Mobius amplitude is 
\ba 
&& \mathcal{M}= -\frac{1}{8} \biggl\{ [(N_o+{\bar N}_o) P_{1}P_{2} + 
(D_{ho}+{\overline{D}_{ho}}) W_1 W_2] (P_{m+1/2} \hat{T}^{B}_{oo}+P_m \hat{T}^{F}_{oo}) \nonumber \\
&& - D_{go} P_1 W_2 [(-1)^n
\hat{T}^{B}_{oo}-\hat{T}^{F}_{oo}] W_3  + D_{fo} W_1 P_2 [(-1)^n
\hat{T}^{B}_{oo}+\hat{T}^{F}_{oo}] W_3  \nonumber\\
&& - \left(\frac{2\hat{\eta}}{\hat{\theta}_{2}}\right)^{2} 
\left[ P_1 ( -(N_o+ {\bar N}_o) \hat{T}^F_{og}+D_{go} \hat{T'}_{og})+
W_1 ( (D_{ho}+ {\bar D}_{ho}) \hat{T}^F_{og}+D_{fo} \hat{T}_{og}) + \right.
\nonumber \\
&& \left. P_2 ( (N_o+ {\bar N}_o) \hat{T}^F_{of}+D_{fo} \hat{T}_{of})  \right. \nonumber \\
&& + W_2 ( -(D_{ho}+ {\bar D}_{ho}) \hat{T}^F_{of}+D_{go} \hat{T'}_{of})
\left. -  (N_o+ {\bar N}_o+D_{ho}+ \overline{ D_{ho}}) (P_{m+1/2}  \hat{T}^B_{oh}- P_m \hat{T}^F_{oh})
\right. \nonumber \\ 
&& \left. + W_3 D_{go} (-\hat{T}^F_{oh}+(-1)^n \hat{T}^B_{oh})-  
W_3 D_{fo} (\hat{T}^F_{oh}+(-1)^n \hat{T}^B_{oh}) \right] \biggr\} \ .  \label{f13}
\ea
The appropriate spacetime interpretation fixes the correct Chan-Paton
parameterization to be 
\ba
&& N_o = n_1 + n_2 + n_3 + n_4 \quad \quad , \quad N_g = n_1 + n_2 - n_3 - n_4
\ , \nonumber \\
&&N_f = i ( n_1 - n_2+ n_3 - n_4) \quad , \quad N_h = i (n_1 - n_2 - n_3 + n_4)
\ , \nonumber \\
&&D_{go} = a + b + c + d \quad \quad  \quad\quad , \quad D_{gg} = a + b - c - d
\ , \nonumber \\
&&D_{gf} = a - b + c - d \quad \quad  \quad\quad , \quad D_{gh} = a - b - c + d
\ , \nonumber \\
&&D_{fo} = o + g + \bar{o} + \bar{g} \ \; \; \quad \quad  \quad , \quad D_{fg} = i ( o - g - \bar{o} + \bar{g})
\ , \nonumber \\
&&D_{ff} = i ( o + g - \bar{o} - \bar{g})\ \   \quad \quad , \quad D_{fh} =  o - g +
\bar{o} - \bar{g}
\ , \nonumber \\
&&D_{ho} = d_1 + d_2 + d_3 + d_4 \quad \    \quad , \quad D_{hg} = i ( d_1 + d_2 -
d_3 - d_4) \ , \nonumber \\
&&D_{hf} = d_1 - d_2 + d_3 - d_4 \quad \   \quad , \quad D_{hh} = -i ( d_1 - d_2 -
d_3 + d_4) \ . \label{f14} 
\ea

The RR tadpole cancellation conditions then read:
\ba 
&&N_o + {\bar N}_o = D_{go} = D_{fo} = D_{ho} + \overline{{D}_{ho}} = 32
 \ , \nonumber \\
&& N_g = \cdots D_{hh}= 0 \ .  \label{f15}
\ea
The spectrum of the model is chiral, with a gauge group
$[U(4)^4]_9 \otimes [USp(8)]^4_{5_1} \otimes[U(8)^2]_{5_2} \otimes
[U(4)^4]_{5_3}$. We worked out in detail the spectrum, but in order to save space, we
display here only the chiral fermionic spectrum of this model. The chiral part of
the spectrum arises in the $D9-D5_{i}$ and the $D5_{i}-D5_{j}$
intersections in the following representations of the appropriate bifundamental
gauge group 
\ba
&&9-5_1 : ({\bf 4,1,1,1;8,1,1,1}) +  ({\bf \bar{4},1,1,1;1,8,1,1})
+ ({\bf 1,4,1,1;1,8,1,1})+
 ({\bf 1,\bar{4},1,1;8,1,1,1}) \nonumber \\ 
&&+ ({\bf 1,1,4,1;1,1,8,1}) +  ({\bf 1,1,\bar{4},1;1,1,1,8}) + ({\bf 1,1,1,4;1,1,1,8})
+ ({\bf 1,1,1,\bar{4};1,1,8,1}) \nonumber \\
&&9-5_2 : ({\bf 4,1,1,1;\bar{8},1}) +({\bf \bar{4},1,1,1;1,8}) +
({\bf 1,4,1,1;1,8}) +({\bf 1,\bar{4},1,1;\bar{8},1}) \nonumber \\
&&+ ({\bf 1,1,4,1;1,\bar{8}}) + ({\bf 1,1,\bar{4},1;8,1}) + ({\bf 1,1,1,4;8,1}) +
 ({\bf 1,1,1,\bar{4};1,\bar{8}}) \nonumber \\
&&9-5_3 :  ({\bf 4,1,1,1;4,1,1,1}) +  ({\bf
  \bar{4},1,1,1;1,1,1,\bar{4}}) + ({\bf 1,4,1,1;1,4,1,1})+
 ({\bf 1,\bar{4},1,1;1,1,\bar{4},1}) \nonumber \\ 
&&+ ({\bf 1,1,4,1;1,1,4,1}) +  ({\bf 1,1,\bar{4},1;1,\bar{4},1,1}) +
({\bf 1,1,1,4;1,1,1,4})
+ ({\bf 1,1,1,\bar{4};\bar{4},1,1,1}) \nonumber \\
&&5_{1}-5_2  : ({\bf 8,1,1,1;\bar{8},1}) +
({\bf 1,8,1,1;1,8}) + ({\bf 1,1,8,1;1,\bar{8}}) + ({\bf 1,1,1,8;8,1}) + \nonumber \\
&&5_{1}-5_{3}  : ({\bf 8,1,1,1;4,1,1,1}) +  ({\bf
  8,1,1,1;1,1,\bar{4},1}) + ({\bf 1,8,1,1;1,4,1,1})+
 ({\bf 1,8,1,1;1,1,1,\bar{4}}) \nonumber \\ 
&&+ ({\bf 1,1,8,1;1,1,4,1}) +  ({\bf 1,1,8,1;\bar{4},1,1,1}) + ({\bf
  1,1,1,8;1,1,1,4}) 
+ ({\bf 1,1,1,8;1,\bar{4},1,1}) \nonumber \\
&&5_{2}-5_3  : ({\bf 8,1;1,1,1,4}) + ({\bf 8,1;1,\bar{4},1,1}) + 
({\bf \bar{8},1;4,1,1,1}) +({\bf \bar{8},1;1,1,\bar{4},1}) \nonumber \\ 
&& + ({\bf 1,8;1,4,1,1}) + ({\bf 1,8;1,1,1,\bar{4}}) +  
 ({\bf 1,\bar{8};1,1,4,1}) + ({\bf 1,\bar{8};\bar{4},1,1,1}) \ . \label{f16}  
\ea

Irreducible gauge anomalies are easily seen to cancel, whereas mixed
anomalies are taken care by the four-dimensional Green-Schwarz
mechanism involving closed sector twisted fields.
The model described  so far has all branes at the origin of the
compact space and therefore coincident $\overline{D5}_1-O5_{1,-}$ and
 $D5_{2}-O5_{2,+}$ systems. In order to make contact with our cosmological solutions, 
we can add Wilson lines on the  $\overline{D5}_1$ and  the $D5_{2}$
branes, in order to get coincident  and BPS system
$\overline{D5}_1-\overline{O5}_{1,+}$ and the non-BPS  system
$D5_{2}-\overline{O5}_{1,-}$, which is precisely the system we are
searching for.


\section{Four dimensional cosmological solutions, nonperturbative
dynamics and moduli stabilization}

In \cite{dmt2} we compactified and T-dualized the 9d time-dependent solution found in 
\cite{dmt1} and we found cosmological solutions for D3 branes, moving
with a constant velocity in a 5d static bulk spacetime, the bulk
being supersymmetric and therefore the metric free of singularities in
the transverse coordinate.  
The five dimensional lagrangian that  describes the T-dual solution with
D3-branes \footnote{We remove prime indices on the T-dual fields,
compared with \cite{dmt2}.} is:
\ba
S_5 &=& {1 \over 2 {\kappa}_5^2} \int d^{5} x \sqrt {-g} \biggl[ 
R^{(5)} - {1 \over 2} ({\partial \Phi})^2 - {40 \over 3} ({\partial
\sigma})^2 - {1 \over 2 \times 5 !} \ e ^{ {40
\sigma \over 3}} \ F_{5}^2 \biggr] \nonumber \\
&-& \int_{X_5=0} d^4 x \biggl[ \sqrt{-\gamma} \ 
 T_0 \ e^{- {20 \sigma \over 3} } \ + q_0 \ A_4 + \cdots
\biggr]  \nonumber \\
&-& \int_{X_5=v_1 X_0} d^4 x \biggl[ \sqrt{-\gamma} \ 
 (T_1 \ e^{- {20 \sigma \over 3}}+ 
{r_c^5 \over g^2} e^{-\Phi} {\rm tr} F^2) \ + q_1 \ A_4 + \cdots
\biggr] \ . \label{n1}
\ea
The classical solution of (\ref{n1}) is \cite{dmt2}
\ba
&&ds_5^2 = [G_0+ {3 |q_0| \kappa^2 \over 2} |X_5|]^{ {2 \over 9}} \ 
[\delta_{\mu \nu} dx^{\mu}  dx^{\nu}- dX_0^2 + d X_5^2 ] \ , \nonumber \\
&&e^{\sigma} = r_c  \ [G_0+ {3 |q_0| \kappa^2 \over 2} |X_5|]^{ {1 \over 6}}
\quad , \quad e^{\Phi} = {\rm const.} \ ,  \label{n2} 
\ea 
with $r_c$ being a constant radius parameter, the coordinates 
$(X_0,X_5)$ being subject to a boost identification
$(X_0 \pm X_5) = \exp (2 \xi) (X_0 \pm X_5)$, with the boost parameter
$\xi$ determined by the mismatch between the tension and the charge of
the non-BPS system \ \cite{dmt1,dmt2} \ $q_1 ch \ \xi = T_1$ as defined in
section 2 and the velocity of the non-BPS system being $v_1 = th \ \xi$. 

By taking appropriate T-dualities, the lower dimensional orbifold
examples we constructed in the Sections 3 and 4 share the same
cosmological solution (\ref{n2}), with the advantage of allowing for nontrivial
dynamics and chiral fermions on the boundary branes. This is due
mostly to the fact that the twisted closed fields in the 6d and 4d
examples of Sections 3,4 have a zero net coupling to the D
branes\footnote{The twisted sector tensors in the 6d model have
actually a physical coupling, with 
opposite signs to the two D9 and/or D5 factors,
which cancel each other and produce no net source in the field eqs.}
and therefore can be 
consistently set to zero in what follows. The compactified
cosmological solution      
of \cite{dmt2} still had flat directions 
and this is most probably phenomenologically unacceptable. Fortunately,
there is a simple argument showing that the classical solution
(\ref{n2}) is valid for a large but finite time evolution. First of
all, it is readily seen from (\ref{n2}) that the
tree-level gauge couplings on the non-BPS system, moving with a
constant velocity in the static bulk
background, are independent of time. This result is corrected by a
one-loop Weyl anomaly. Indeed, one of the steps undertaken in
\cite{dmt2} was a Weyl rescaling in 5d from the string frame to the
Einstein frame:
\be
g^{(5)}_{\alpha \beta} \quad = \quad  
\exp ({{\Phi \over 2} - {10 \sigma \over 3}})
\ g^{(5)}_{E, \alpha \beta} \ , \label{n01}  
\ee
where we remind the reader that we removed in the present paper the
primes on the T-dual fields compared to \cite{dmt2}. This Weyl
transformation is seen by the D3 brane fields as a 4d Weyl
transformation, which is subject to the standard one-loop anomaly. The
Weyl anomaly corrected gauge couplings become
\be
{1 \over g_{YM}^2} \quad = \quad \ {r_c^5 \over g^2} \ e^{-\Phi} + 
({20 \sigma \over 3} - \Phi) \ {b_1 \over 32 \pi^2} \ . \label{n02} 
\ee 
Since, by using (\ref{n2}) on the non-BPS system trajectory$X_5 = v_1 X_0$, the breathing mode
$\sigma$ is increasing in time, this reflects in an
effective logarithmic (Einstein proper) time dependence of the gauge couplings   
\be
{1 \over g_{YM}^2 (\tau_E)} \quad = \quad {1 \over g_{YM,0}^2} + 
{b_1 \over 32 \pi^2} \ \ln ({\tau_E \over \tau_{E,0}}) \ . \label{n03} 
\ee 
It is interesting to notice the analogy of (\ref{n03}) with the one-loop renormalization group
equations, but with energy replace by time. In the toroidal
compactified model and a large class of other examples,
the non-BPS system has a matter content corresponding to an
asymptotically-free gauge group $b_1 < 0$. This means that, after a
(exponentially) long time evolution, the non-BPS system enters a
nonperturbative regime which must stop the time evolution or, in any
case, must invalid the simple solution (\ref{n2}) combined with the
constant velocity trajectory $X_5 = v_1 X_0$. 
The logarithmic nature of the time evolution is crucial
for the cosmological realization of the hierarchies discussed in
\cite{dmt2}. Indeed, if the time dependence was a power-law, the gauge
coupling was either going very fast to zero (for $b_1 >0$), either the system was 
entering, within a time evolution of the order of the Planck time,
a nonperturbative regime, before generating any interesting
physical effect. We believe (but we have no proof of this statement)
that the tree-level bulk supersymmetry plays a crucial role in
generating the desired logarithmic time dependence.    
The time $\tau_{\rm max}$ where this happens and the corresponding
maximal value of the radii of the internal five-torus are
\be
\tau_{\rm max} = \tau_{E,0} \ \exp \{ {32 \pi^2 \over |b_1| \ g_{YM,0}^2} \} \quad , \quad
e^{\sigma_{\rm max}} = R_c \ \exp \{ {24 \pi^2 \over 5 |b_1| \ g_{YM,0}^2} \} \ , \label{n04}
\ee    
with $R_c$ the initial value of the compact radii. As seen in
(\ref{n04}), the internal radius becomes exponentially large before
entering the nonperturbative regime on the non-BPS system,
supporting the perturbative nature of the hierarchies proposed in \cite{dmt2}.
Any mechanism of moduli stabilization, in order to be viable, must
produce a value of the compact radius smaller than its maximally
admitted one (\ref{n04}). Going one step further, by adding the
standard one-loop RGE contribution to find the full
one-loop (energy-dependent) gauge couplings, we find
\be
{1 \over g_{YM}^2 (\mu)} \quad = \quad \ {r_c^5 \over g^2} \ e^{-\Phi} + 
 {b_1 \over 8 \pi^2} \ \ln \ \biggl({\Lambda_0 \exp (-\Phi/4+ {5
\sigma / 3}) \over \mu } \biggr)  \ , \label{n05} 
\ee    
where $\Lambda_0$ is the (string scale) UV cutoff. Notice that the
effect of the time dependence can be absorbed into a time growth 
of the UV cutoff. The scalar
potential induced by the gaugino condensation can be estimated, as
usual, as the fourth power of the energy scale where couplings become
strong
\be
V \ \sim \ \Lambda^4 \ = \ \Lambda_0^4 \ \exp ({{20 \sigma \over 3} - \Phi})   
\ \exp \{- {32 \pi^2 \over |b_1| \ g_{YM,0}^2} \} \ \equiv \ \alpha \ \exp ({20
\sigma \over 3}) \ , \label{n06}
\ee 
where 
\be
\alpha \ \equiv \ \Lambda_0^4  \ \exp (- \Phi_0) \ \exp \{- {32 \pi^2 \over
|b_1| \ g_{YM,0}^2} \} \ . \label{n07}
\ee 
In the last step, we anticipated
that a constant dilaton is a solution of the brane dynamics
(for example, as we will discuss in more details later on, gaugino
condensation combined with additional NS-NS fluxes \cite{din}), 
whereas by an appropriate combination of NS-NS and RR
fluxes \cite{gkp} it can be stabilized. 
An heuristic argument suggests that the nonperturbative potential  
(\ref{n06}) can stabilize the radius field $\sigma$. Indeed,
considering the already existing potential (coming from the tension)
and adding it to (\ref{n06}), we find
\be
V_{\rm tot}  \quad = \quad T_1 \ \exp (-{20 \sigma \over 3}) \ + \ 
\alpha_1 \ \exp ({20 \sigma \over 3}) \ , \label{n08}
\ee
which has a minimum of the order $\exp ({40 \sigma_0 /3}) \sim (T_1 /
\alpha_1) >>1$. By using (\ref{n07}), we learn that this is an
exponentially large value, as required for generating hierarchies
advocated in \cite{dmt2}. It is therefore very important that $\alpha$ is much
smaller than its naive value $\alpha << \Lambda_0^4 $.  Moreover,  
$ \sigma_0 << \sigma_{\rm max}$ and therefore the stabilization
procedure is under control. The more
detailed study in the following shows that the correct value of
$\sigma_0$, obtained by solving the field equations, has indeed the order of
magnitude of the minimum coming from (\ref{n08}), but we actually need
additional terms in the scalar potential combining with the gaugino
condensation into a perfect square.  The resulting numerical
coefficient will be slightly different as well. 
  
In order to stabilize moduli we need to look more closely on the
possible nonperturbative dynamics. There are two possibilities that we
will consider in the following~:

\noindent (i) Brane potentials.

Nonperturbative effects like gaugino condensation \cite{din} on D-branes or
loop perturbative effects generated by supersymmetry breaking
on some D-branes naturally generate potentials
for the closed (bulk) fields localized on the D-branes. 
They were already invoked some time ago in a phenomenological approach by Goldberger and
Wise \cite{gw} in connection with moduli stabilization. In our case,
we argued that these nonperturbative effects are naturally triggered
by the time evolution. The induced 
potentials, called $V_i$ in what follows, where $i = 0,1$ index the two spacetime
boundaries, change  our cosmological solutions 
in a way that is the main concern of this section.

\noindent (ii) NS-NS and RR fluxes.     

Our perturbative orientifold models containing (after T-dualities) 
D3/O3 and D7/O7 branes and O-planes allow for the introduction of
NS-NS and RR fluxes, along the lines of \cite{gkp}. Indeed, while the
closed string spectrum and the orientifold projection in the vacua we
considered are different compared to the simple IIB orientifold
considered in \cite{gkp}, the massless untwisted spectrum contains the
same RR and NS-NS fields. The additional left world-sheet fermion number in
our orientifold projection has a trivial action on them and the
analysis of possible fluxes to add is similar to \cite{gkp} and, in
context of supersymmetric orbifold cousins of the ones we discuss,
was performed in \cite{blt}. Fluxes have 
the effect of generating potentials (which are however not
brane-localized) for closed string fields \cite{gvw} and to stabilize
the dilaton and the complex structure moduli.

In order to stabilize moduli fields we add therefore
potentials on the boundaries generated by nonperturbative and/or
perturbative effects :

\ba 
S_V= -\int_{y=0}d^4 x \sqrt{-\gamma}\ V_0 (\sigma, \Phi) - 
\int_{y=y_1} d^4 x \sqrt{-\gamma}\ V_1(\sigma, \Phi) \ ,  \label{n09} 
\ea

where in what follows we will mostly consider (\ref{n07}), (\ref{n08}). 
As we will see, in the cases we are discussing, (\ref{n09}) does not
stabilize the dilaton, which is still a flat direction of the full
effective action. whereas suitable fluxes can stabilize it.

Appropriate brane
potentials will stabilize the $\sigma$ and $y_1$ moduli fields. The non-BPS brane, which was
moving into a noncompact space subject to a boost identification
\cite{dmt1,dmt2}, will stop moving. Stabilization of the distance
between the two boundary branes means that the coordinate $X_5 $ becomes
compact again and will be denoted by $y$ in the following. The
stabilization consists in our case in finding explicit solutions to
the field eqs. for the lagrangian (\ref{n1})+(\ref{n09}) and imposing
appropriate boundary conditions at the position of the boundary branes
$y=0$ and $y = y_1$.

We are in what follows searching for solutions of the field
equations of the form:

\be ds_5^2 = e^{2A(y)} \ g_{\mu\nu}dx^{\mu}dx^{\nu} + e^{2B(y)} \ dy^2 \ , \label{n3}
\ee

with $g_{\mu\nu}$ being the Minkowski metric, $\eta_{\mu\nu}$, or the de
Sitter metric, $diag(-1,e^{2\sqrt{\Lambda}t}\delta_{ij})$.

\subsection{Minkowski solution}

We are using the ansatz :
\ba 
ds_5^2 &=& e^{2A(y)} \ \eta_{\mu\nu}dx^{\mu}dx^{\nu} + e^{2B(y)} \
dy^2 \ , \nonumber \\
F_5 &=& \tilde{f}(y) \ \epsilon_{5} \quad , \quad \sigma=\sigma(y) \ , 
\quad \Phi = \Phi_0 = {\rm const.} \ , \label{m1}
\ea

where $\epsilon_{5}$ is the five-dimensional volume form. The equation
of motion of the five form has the solution:

\be
\tilde{f}=-q_0 \ k_5^2 \ e^{4A+B-{40\over 3}\sigma} \ \epsilon (y) \ , \label{5f}
\ee
 where $\epsilon(y)$ is an odd $2y_1$-periodic function  
and $\epsilon(y)=1$ when $y$ is between $0$ and $y_1$ .

 Replacing this solution 
in the Einstein, $\sigma$ and $\Phi$ equations, we obtain\footnote{The
notation we use is $'=\frac{d}{dy}$.} :

\ba
&& 3A''+6A'^2-3A'B'+\frac{20}{3}\sigma'^2+\frac{1}{4}(k_5^2q_0)^2e^{2B-\frac{40}{3}\sigma}
= -k_5^2e^{B-\frac{20}{3}\sigma}\biggl[T_0\delta(y)+T_1\delta(y-y_1)\biggr]-\nonumber\\
&&- k_5^2 e^B\biggl[V_0 \delta(y)+V_1 \delta(y-y_1)\biggr] \ ,
\nonumber \\
&&
6A'^2-\frac{20}{3}\sigma'^2+\frac{1}{4}(k_5^2q_0)^2e^{2B-\frac{40}{3}\sigma}=0
\ , \nonumber \\
&& \sigma''+4A'\sigma'-B'\sigma'+{1\over 4}(k_5^2q_0)^2e^{2B-\frac{40}{3}\sigma}
= -{k_5^2 \over 2} e^{B-{20\over 3}\sigma}\biggl[T_0\delta(y)+T_1\delta(y-y_1)\biggr]+\nonumber\\
&& +{3\over 40}k_5^2e^B\biggl[\delta(y)\frac{\partial
V_0}{\partial \sigma} + \delta(y-y_1)\frac{\partial
V_1}{\partial \sigma}\biggr] \ , \nonumber \\
&& \Phi^{''} + 4 A' \Phi^{'} = 2 k_5^2 e^B\biggl[\delta(y)\frac{\partial
V_0}{\partial \Phi} + \delta(y-y_1)\frac{\partial V_1}{\partial
\Phi}\biggr] \ . 
 \label{sigma}
\ea

By a change of coordinate $y$ we can fix the gauge to $B=20\,\sigma/3 $.
The equation (\ref{sigma}) shows that $A'$ and $\sigma'$ can be
parametrized by a function $f$ as follows :

\be 
A'=\pm {z\over\sqrt{6}}\ sh f\ \epsilon(y) \quad , \quad
\sigma' =  \pm z \sqrt{3\over 20}\ ch f \  \epsilon(y) \ , \label{m01} 
\ee

where $z={1\over 2}k_5^2 |q_0|$. 

The $(+,-)$ and $(-,-)$ solutions are incompatible with the boundary
conditions below and will not be discussed anymore.
The bulk part of the two remaining equations in (\ref{sigma}) is the
same, as it should, and takes the form:

\be 
\sqrt{3} f' \pm 2\sqrt{2}z \ ch f - 2\sqrt{5} \ z \ sh f=0 \ , \label{m2}
\ee
where the $+$ sign in (\ref{m2}) corresponds to the $(+,+)$ case
whereas the $-$ sign corresponds to the case $(-,+)$ in (\ref{m01}).
 
The solutions of (\ref{m2}) with $+$ sign are :
\be  e^f= -a \ th (z|y|+C) \quad , \quad  {\rm and} \quad  
e^f= -a \ cth (z|y|+C) \ , \label{m03}
\ee
whereas the ones with $-$ sign are 
\ba  
&& e^f = -  a ^{-1}\ th (z|y|+C) \quad , \quad  {\rm and} \quad  e^f=
-a^{-1} \ cth (z|y|+C) \ , \nonumber \\
&& {\rm where} \quad a = \frac{\sqrt{5}+\sqrt{2}}{\sqrt{3}} \  \label{m02}
\ea
and $C$ is an integration constant. 



Next we have to impose the boundary conditions given by the boundary D-branes/O-planes
\ba 
\sqrt{3\over 2}\  |q_0| \ sh f &=& \mp  \rl T_0+e^{\frac{20\sigma}{3}}V_0\rr
|_{y=0} \ , \nonumber \\
\sqrt{3\over 5} \ |q_0| \ ch f &=& - \rl T_0-{3\over 20}
e^{\frac{20\sigma}{3}}\frac{\partial V_0}{\partial \sigma}\rr|_{y=0} \
, \nonumber\\
\frac{\partial V_0}{\partial \Phi}|_{y=0} &=& 0 \quad , \quad
\nonumber \\
\sqrt{3\over 2}\  |q_0| \ sh f &=& \pm  \rl T_1+e^{\frac{20\sigma}{3}}V_1\rr
|_{y=y_1} \ , \nonumber\\
\sqrt{3\over 5} \ |q_0| \ ch f &=& + \rl T_1-{3\over 20}
e^{\frac{20\sigma}{3}}\frac{\partial V_1}{\partial \sigma}\rr|_{y=y_1}
\ , \nonumber \\
\frac{\partial V_1}{\partial \Phi}|_{y=y_1} &=& 0 \quad ,  
\label{m4}
\ea

 where the first sign refers to the $(+,+)$ case, while the second sign 
refers to the $(-,+)$ case.

The simplest and actually the only solution we were able to find
compatible with boundary conditions (\ref{m4}), for brane potentials of
physical interest is, for both $(+,+)$ and $(-,+)$
cases, $C=-\infty$, so that $e^f$ is constant
 
\be e^{f(y)}=a \quad  {\rm for} \quad (+,+) \quad ; \quad
    e^{f(y)}=a^{-1}\quad {\rm for} \quad (-,+) \ . \label{m04}
\ee

Notice that $ch f = \sqrt{5 /3}$ and $sh f = \pm \sqrt{2 /3}$ and
consequently the boundary conditions take a particularly simple form.
Both cases give the same solution for $A$ and $\sigma$
\be 
A(y)={z\over 3}|y|+C_A \quad , \quad \sigma (y)={z\over
2}|y|+C_{\sigma} \label{sol}
\ee 

and the metric takes the form:
\be 
ds_5^2 \quad = \quad e^{{2\over 3}z|y|} \ \eta_{\mu\nu}dx^\mu dx^\nu+e^{{20 \over
3}z |y|+{40\over 3} C_\sigma} \ dy^2 \ , \label{m05}
\ee

where $C_A$ was absorbed  by a rescaling of the $x^{\mu}$
coordinates. By a change of coordinate in $y$, it  turns out that
(\ref{m05}) is the same as (\ref{n2}), which is the T-dual to the one
worked out in \cite{pw}.  This result reflects the
(approximate) supersymmetry of the bulk space at the lowest order in
perturbation theory. 
In this case, combining in a straightforward way the
boundary conditions (\ref{m04}) and defining the total brane scalar
potentials (including the tension contribution) and their sum
\ba
&&V_{i, {\rm tot}} \equiv V_i (\phi_a) \ + \ T_i \ e^{-{20 \sigma \over 3}} \ , \nonumber \\
&& U (y) \ = \ V_{0, {\rm tot}} \ \delta (y) \ + \ V_{1, {\rm tot}} \ \delta (y-y_1) \
, \label{m06} 
\ea  
we find the local conditions
\be
<V_{i, {\rm tot}}> \ = \ <V_{SUSY}>|_{y=y_i} \quad , \quad 
<{\partial V_{i, {\rm tot}} \over \partial \phi_a}> \ = \ <{\partial V_{SUSY}
  \over \partial \phi_a}>|_{y=y_i} \ , \label{m07}
\ee

where $\phi_a = \sigma,\Phi$ and where $V_{SUSY} = q_i \exp (-20 \sigma / 3)$ is the BPS
(supersymmetric) tension potential. From now on, in order to avoid
confusion, we use the notation $<f
(\phi_a)>|_{y=y_i}$ to denote the numerical value of the function $f
(\phi_a)$ evaluated by inserting the classical solution $\phi_a = \phi_a (y_i)$ of the field
equations (\ref{sigma}) with the boundary conditions (\ref{m4}).

The conditions (\ref{m07}) have an
obvious interpretation~: since the bulk is supersymmetric to lowest
order, branes sources (the potential and its derivatives) should mimic exactly 
the supersymmetric case. Integrating over the compact coordinate we
find the four dimensional integrability conditions
\be
\int_0^{y_1} dy\ \sqrt{g_{yy}}\ <U> \ = \ 0 \quad , \quad
\int_0^{y_1}dy\ \sqrt{g_{yy}}\ < {\partial U \over
  \partial \phi_a}> \ = \ 0 \ . \label{m08}
\ee
By defining the four dimensional potential 
\be
V_4 \ = \ \int_0^{y_1} dy\ \sqrt{g_{yy}}\ U \ = \ 
(\sqrt{g_{yy}} V_{0, {\rm tot}})|_{y=0} \ + \ 
(\sqrt{g_{yy}} V_{1, {\rm tot}})|_{y=y_1} \ , 
\ee 
we find the transparent
four-dimensional conditions
\be
<V_4> = 0 \quad , \quad  <{\partial V_4 \over  \partial \phi_a}> = 0 \ , \label{m09}
\ee
where in 4d the $<f>$ symbol has now the standard interpretation of
evaluating the function $f$ in the vacuum of the 4d theory, obtained
by minimizing the 4d potential $V_4$. 
The second  equation (\ref{m09}) defines the minima of the
potential, while the first reminds us that we are searching for a
Minkowski solution and therefore the 4d cosmological constant is zero
in the vacuum.  


Since in $y=0$ we have a BPS brane, the simplest and most natural solution is $V_0=0$ and this is
the case we are considering to start with. 

Let us consider concrete examples of scalar potentials, of the form
\be
V_1 \quad =\quad \alpha_1 \ e^{\beta_1 \sigma} \ + \ 
\alpha_2 \ e^{\beta_2 \sigma} \ . \label{m020}
\ee 
The boundary conditions impose the relation:

\be 
\alpha _1 \ (1+{3\over 20}\beta_1) <e^{\beta_1\sigma}>\ |_{y=y_1} +
\alpha_2 \ (1+{3 \over 20} \beta_2) <e^{\beta_2\sigma}>\ |_{y=y_1}=0 \ . \label{m010} 
\ee

In the following we are discussing 2 cases :

(i) $\alpha_2=0$ . In this case the boundary conditions in $y=y_1$ imply:

\be 
\beta_1=-{20\over 3} \quad {\rm and} \quad \alpha_1=|q_0|-T_1 < 0  \ . \label{m011}
\ee

The total potential becomes :

\be 
V_{1,{\rm tot}} \ = \ T_1 e^{-{20\over 3}\sigma}+\alpha_1 e^{-{20\over
    3}\sigma} \ = \ q_1 \ e^{-{20\over 3}\sigma} \ . \label{m012}
\ee

 We recover the supersymmetric case, with a BPS brane in $y=y_1$ and
$y_1$ that is not stabilized.  Notice that
$<{\partial V_{1,{\rm tot}}\over \partial \phi}>=0$ automatically, and
therefore all equations of motion and boundary conditions are
satisfied in this case. We can now add, in order to stabilize the
dilaton, NS-NS and RR fluxes. 

(ii) $\beta_1={20\over 3}\ , \ \beta_2=0$. The boundary conditions
imply :

\be 
\alpha_1 \ <e^{{40\over 3}\sigma}>\ |_{y=y_1}=T_1-|q_0|>0 \quad  {\rm and} \quad
 \alpha_2=-2\sqrt{\alpha_1(T_1-|q_0|)}<0\label{alpha1} \ . \label{m013}
\ee
Notice, by using  (\ref{n07}) $\alpha_1 <<1$, that from (\ref{m013})
the volume modulus $\sigma$ is stabilized to a very large value,
qualitatively of the order of magnitude given by the naive argument
with the (incomplete) scalar potential (\ref{n08}). By inserting into
(\ref{m013}) the classical solution (\ref{sol}) we find
\be
e^{{20 z y_1 \over 3}+ {40 C_{\sigma} \over 3}} = {T_1 - q_1 \over
  \alpha_1} >> 1 \ . \label{y1}
\ee
We see that $y_1$ can be stabilized to a moderately large value and
therefore it creates no potential phenomenological problems like
deviations from the gravitational attraction at macroscopic distances. 
Actually only a linear combination of $\sigma$ and $y_1$ is stabilized
by (\ref{y1}). In order to separately stabilize $\sigma$ and $y_1$, 
nontrivial dynamics in $y=0$ seems to be necessary.
 
 The total potential reads now :

\be 
V_{1, {\rm tot}}=\rl \sqrt{T_1-|q_0|}e^{-{10\over
3}\sigma}-\sqrt{\alpha_1}e^{{10\over 3}\sigma}\rr^2+q_1e^{-{20\over
3}\sigma} \ . \label{m014}
\ee

 By using (\ref{alpha1}), the positive (squared) term vanishes
 evaluated as a solution of the classical field equations and the
 (expectation value of the) scalar potential and its derivative mimic, as is required by
(\ref{m07}), the supersymmetric potential. As a consequence,
 $<{\partial V_{1,{\rm tot}}\over \partial \phi}>|_{y=y_1}=0$ is
automatically satisfied, so again
a constant dilaton (eventually stabilized by fluxes) is still a
solution after adding the induced brane potential. The four
dimensional scalar potential in this case is
\be
V_4 = \bigl(\sqrt{T_1-|q_0|} -\sqrt{\alpha_1} e^{{20\over 3}\sigma} \bigr)^2  \ 
\ee
and is positive definite, like in supersymmetric theories. Before
adding fluxes, we find therefore results very similar to the no-scale
supergravity models \cite{cfkn}: positive definite scalar potential and one flat
direction (the dilaton). This is intriguing and is, presumably,
related to two facts. First of all , the bulk being almost
 supersymmetric and coupling to the non-BPS system in $y = y_1$, the full effective lagrangian 
has a non-linearly realized supersymmetry on the non-BPS system. 
The condition of having static solutions and bulk supersymmetry force
the brane scalar potentials to have a form similar to the standard supergravity
lagrangian which, in all effective string models, is of the no-scale
type. Therefore, despite the non-BPS system in $y = y_1$, the
dynamics responsible for the stabilization of the fields $\sigma$ and
$y_1$, which describe Kahler moduli in string language, respect
constraints very similar to models with spontaneously broken
supersymmetry. This point clearly deserves, in our opinion,  further and
more detailed studies.

We can add, consistently with the boundary conditions (\ref{m4}),
induced brane potentials on the $y=0$ brane, in order to stabilize both
$\sigma$ and $y_1$ moduli. The simplest example doing the job is the
positive potential $V_0 = (\gamma_1 \exp(\delta_1 \sigma) +\gamma_2
\exp(\delta_2 \sigma))^2 $. In this case $\sigma$ is stabilized via the
condition $\gamma_1 \exp(\delta_1 C_{\sigma}) +\gamma_2
\exp(\delta_2 C_{\sigma})=0$ and then $y_1$ is stabilized
via (\ref{y1}). This example is similar to the racetrack examples of
heterotic dilaton stabilization \cite{racetrack}.  
In analogy
with the previous example, $V_0$ can, via $\gamma_i$, depend on the
dilaton, which is, however, still not stabilized by the $y=0$ brane dynamics.

\subsection{de Sitter solution}

We are looking for solutions of the form:

\ba ds_5^2&=&e^{2A(y)+2C(t)}(-dt^2+\delta_{ij}dx^idx^j)+e^{2B(y)}dy^2\ \ ,
\nonumber \\
F_5 &=& \tilde{f}(t,y) \ \epsilon_{5} , \quad \sigma=\sigma(y) \
, \quad \Phi = \Phi (y) \ . 
\label{m12}
\ea

As before we can easily determine the solution for the five form from
its equation of motion:

\be 
\tilde{f}=-q_0 \ k_5^2 e^{4A+B+4C-{40\over 3}\sigma} \ \epsilon (y) \ .
\ee

 Fixing the gauge to $B=20\sigma/3$ we are left with the equations\footnote{The
notation we use is $'=\frac{d}{dy}$ and $\dot{}=\frac{d}{dt}$.}:
\ba &&-(2\ddot{C}+\dot{C}^2)e^{-2A-2C+{40\sigma \over 3}}+
3A''+6A'^2-20A'\sigma'+{20\over 3}\sigma'^2+{1\over 4}(q_0k_5^2)^2=\nonumber\\
&& = -k_5^2\biggl[T_0\delta(y)+T_1\delta(y-y_1)\biggr]
- k_5^2 e^{20\sigma \over
3}\biggl[V_0\delta(y)+V_1\delta(y-y_1)\biggr]\label{munus} \ , 
\nonumber \\
&&3\dot{C}^2e^{-2A-2C+{40\sigma \over
3}}-3A''-6A'^2+20A'\sigma'-{20\over 3}\sigma'^2-{1\over
4}(q_0k_5^2)^2=\nonumber\\
 &&=k_5^2\biggl[T_0\delta(y)+T_1\delta(y-y_1)\biggr]
+ k_5^2 e^{20\sigma \over
3}\biggl[V_0\delta(y)+V_1\delta(y-y_1)\biggr] \ , \label{00s} \nonumber\\
&&-3(\ddot{C}+\dot{C}^2)e^{-2A-2C+{40\sigma \over 3}}+6A'^2-{20\over 3}\sigma'^2+
{1\over 4}(q_0k_5^2)^2= 0 \ , \label{55s} \nonumber \\
&&\sigma''+4A'\sigma'-{20\over 3}\sigma'^2+{1\over
4}(q_0k_5^2)^2=\nonumber\\
&&=-{k_5^2\over
2}\biggl[T_0\delta(y)+T_1\delta(y-y_1)\biggr]+{3\over 40}k_5^2e^{20\sigma\over 3}\biggl[\delta(y)\frac{\partial
V_0}{\partial \sigma} + \delta(y-y_1)\frac{\partial
V_1}{\partial \sigma}\biggr]\nonumber \\
&& \Phi''+\Phi'\rl 4A'-{20\over 3}\sigma'\rr=0 \ . \label{sigmas}
\ea

 The first two equations determine the function $C(t)$ which, after
shifting the origin of time, is given by :
\be \ddot{C}=\dot{C}^2 \quad \Rightarrow \quad e^{2C}=\frac{1}{C_1^2
t^2} \  \label{m13}
\ee

and therefore (\ref{m12}) describes a warped 4d de Sitter metric. 

 If the cosmological constant, $\Lambda=C_1$, is small we can
 look for solutions that are
 small perturbations around the Minkowski solution:
\be A=A_0+\tilde{A} \quad  , \quad \quad
 \sigma = \sigma_0 + \tilde{\sigma} \quad  , \quad \quad
\Phi=\Phi_0+\tilde{\phi} \ , \label{m14} 
\ee

where $A_0$ and $\sigma_0$ are the Minkowski solutions and $\tilde{A}$
and $\tilde{\sigma}$ are small perturbations.

 The linearized field equations take the form:

\ba 
&&3\tilde{A}''+\tilde{A}'(12
A_0'-20\sigma_0')+\tilde{\sigma}\rl{40\over
  3}\sigma_0'-20A_0'\rr-3C_1^2e^{-2A_0+{40\over
    3}\sigma_0}=0 \ , \nonumber\\
&&\tilde{\sigma}''+4\tilde{A}'\sigma_0'+\tilde{\sigma}'\rl
4A_0'-{40\over 3}\sigma_0'\rr=0 \ , \nonumber\\
&&12\tilde{A}'A_0'-{40\over
3}\tilde{\sigma}'\sigma_0'-6C_1^2e^{-2A_0+{40 \over
3} \sigma_0}=0\nonumber\\
&&\tilde{\phi}''+\tilde{\phi}'\rl 4A_0-{20\over 3}\sigma_0 \rr=0 \
. \label{m15} 
\ea
Strictly speaking there are also 4d localized sources in
(\ref{m15}). The scalar potentials (\ref{n07})-(\ref{n08}) inspired
by the gaugino condensation are themselves small. The sources in
(\ref{m15}) are proportional to them and also to the small perturbations 
and are therefore quadratically small.  
  
The last two equations allow to determine a first order equation for
the variable $\tilde{\sigma}'=g$:

\be 
g'+4 g \rl A_0'-{10\over 3}\sigma_0'+{10\over 9}{\sigma_0'^2\over
  A_0'}\rr = -2 {\sigma_0'\over A_0'}C_1^2e^{-2A_0+{40\over
3}\sigma_0} \ . \label{m16}
\ee

Looking for solution of the form $g=f\chi$, with $f$ the solution of
the homogeneous equation we find:
\be
f = e^{-4\int\rl  A_0'-{10\over 3}\sigma_0'+{10\over 9}{\sigma_0'^2\over
  A_0'}\rr dy \ +C_f} \ , \
\chi = - 2 C_1^2 \int{1\over f}{\sigma_0'\over A_0'}\ e^{-2A_0+{40\over
    3}\sigma_0} dy +C_\chi\ . \label{m17}
\ee
We showed therefore that de Sitter solutions exists, at least in the
vicinity of Minkowski one. In order to get a small four-dimensional
cosmological constant, we need basically the same fine-tuning as in the
Minkowski solutions. 

 Using the Minkowski solution found in Section 5.1 the linearized
field equations (\ref{m15}) become:

\ba 
&&\tilde{A}'' -2z \ \tilde{A}'= C_1^2\ e^{6zy-2C_A+{40\over
    3}C_{\sigma}}\ , \nonumber\\
&&\tilde{\sigma}''+2z\ \tilde{A}'-{16\over 3}\ \tilde{\sigma}'=0 \ ,
\nonumber\\ 
&&2z\ \tilde{A}'-{10\over
3}\ \tilde{\sigma}'=3C_1^2\ e^{6zy-2C_A+{40\over
    3}C_{\sigma}}\ ,\nonumber\\ 
&&\tilde{\phi}''+\tilde{\phi}'\rl -2zy+4C_A-{20\over 3}C_{\sigma}\rr=0 \ . 
\ea

 Finally the metric and dilaton read:

\ba ds_5^2&=&e^{{2\over 3}z|y|+{C_0\over
z}e^{2z|y|}+C_{A\sigma}{C_1^2\over 12  z^2}e^{6z|y|}}[ {1\over
C_1^2t^2}(-dt^2+\delta _{ij}dx^i dx^j)+\nonumber\\
&+&C_B\ e^{6z|y|+{C_0\over
z}e^{2z|y|}-{7\over 4}C_{A\sigma}{C_1^2\over
z^2}e^{6z|y|}}dy^2]\ ,\nonumber\\
\Phi&=&\Phi_0+{C\over 2z}e^{2z|y|}+C_{\phi}\ .\ea

It is likely, but we didn't check it, that in a different range of the
parameters there are also 4d anti de Sitter solutions. The main point of
all our discussions on nonpertubative dynamics versus moduli
stabilization is that, due to the smallness of the nonperturbative
scalar potentials, moduli are stabilized at large values by allowing
hierarchies to be generated.
    
\section{Conclusions}

String models with broken supersymmetry are the natural candidates to
study cosmology and moduli stabilization. We showed that non-tachyonic
orbifold string vacua in various dimensions can be constructed, 
starting from Scherk-Schwarz compactifications in the closed sector,
with the would-be tachyon removed by an appropriate orientifold projection
which generates a peculiar $Op_{+}-\overline{Op}_{-}$ orientifold plane
structure. Including by consistency the open sector, we find
$Dp-\overline{Op}_{-}$ systems which are very similar to the
brane-antibrane systems, that were shown by A.Sen \cite{sen} 
to have interesting cosmological applications, but are now tachyon-free,
allowing for a clean classical description of their dynamics.

Our non-tachyonic and non-BPS vacua generate
simple time-dependent solutions corresponding to  (one or some of the) 
spacetime boundaries moving with a constant velocity in a static bulk,
subject classically to a boost identification. Gauge (and Yukawa) couplings  
on the D3 branes are time-independent at tree-level and acquire a
logarithmic time-dependence (\ref{n03}) very similar to that induced by the standard RG
equations. For asymptotically free gauge groups, which are generically
obtained in the non-BPS configurations $D3-\overline{O3}_{-}$ of our
string vacua, after an exponentially (in fundamental string units) long
time (\ref{n04}) the non-BPS branes enter a nonperturbative regime and 
fix part of the scalar moduli, including the fields describing 
the size of compact space, but not the (T-dual) dilaton. This
exponentially long time can
generate hierarchies between the fundamental string scale and the
Planck scale.  
In order to stabilize the internal radii,
which are Kahler moduli in string language,  we showed that
stabilization potential must produce exactly the same sources (vev's
of the potentials and their first derivatives) as in the supersymmetric  
situation. We started the study of the ``inverse problem'' of
determining additional brane potentials compatible with static solutions in the
non-BPS vacua. In the particular case of gaugino condensation on the
non-BPS boundary, 
we need to combine the condensate with a constant potential term into a
positive definite scalar potential, in close analogy with the
dilaton stabilization in the heterotic string \cite{din}. Contrary to
the heterotic example, however, in our case
the Kahler moduli are fixed, whereas the dilaton remains a
flat direction that can subsequently be stabilized by adding
appropriate fluxes. 
After stabilization, the classical solution becomes exactly the same as
the supersymmetric one, nonperturbative brane dynamics ``repairing''   
the non-BPS nature of brane sources in order to match the
corresponding BPS one.  

The exponentially long time period of validity of our cosmological 
solutions and the subsequent nonperturbative phenomena triggering moduli 
stabilization raise the hope of possible imprints of string physics in
early cosmology. From this perspective, an analysis along the lines
of \cite{greene} of the class of models we constructed in the present
paper would be very interesting and useful to perform.

\noindent
{\bf Acknowledgments.} 
We thanks Jihad Mourad and Augusto Sagnotti for useful discussions. The work of E.D. was 
supported in part by the RTN European Program
HPRN-CT-2000-00148. E.D. is grateful for the warm hospitality at
TPI-Minnesota and the Theory Division of CERN at early stages of this work. 
 

\section{Appendix : Characters for $Z_2 \times Z_2$ orbifolds}

Our conventions for writing partition functions and defining moduli
for various string surfaces are the ones in the third and the fourth
reference in \cite{reviews}. 
The level one SO(2n) characters are:
\ba \label{a1}
O_{2n}&=&\frac{\theta_{3}^n+\theta_{4}^n}{2 \eta^n}~,\qquad ~
V_{2n}=\frac{\theta_{3}^n-\theta_{4}^n}{2 \eta^n}~,\nonumber \\
S_{2n}&=&\frac{\theta_{2}^n+i^n \theta_{1}^n}{2 \eta^n}~,\qquad
C_{2n}=\frac{\theta_{2}^n-i^n \theta_{1}^n}{2 \eta^n}~ . 
\ea
The $Z_2 \times Z_2$ torus partition function in the text is written in
terms of the following supersymmetric characters \cite{bianchi,aads}:
\ba \label{a2}
\tau_{oo}&=& V_{2}O_{2}O_{2}O_{2} + O_{2}V_{2}V_{2}V_{2} - S_{2}S_{2}S_{2}S_{2} - C_{2}C_{2}C_{2}C_{2}\,\nonumber\\
\tau_{og}&=& O_{2}V_{2}O_{2}O_{2} + V_{2}O_{2}V_{2}V_{2} - C_{2}C_{2}S_{2}S_{2} - S_{2}S_{2}C_{2}C_{2}\,\nonumber\\
\tau_{oh}&=& O_{2}O_{2}O_{2}V_{2} + V_{2}V_{2}V_{2}O_{2} - C_{2}S_{2}S_{2}C_{2} - S_{2}C_{2}C_{2}S_{2}\,\nonumber\\
\tau_{of}&=& O_{2}O_{2}V_{2}O_{2} + V_{2}V_{2}O_{2}V_{2} - C_{2}S_{2}C_{2}S_{2} - S_{2}C_{2}S_{2}C_{2}\,\nonumber\\
\tau_{go}&=& V_{2}O_{2}S_{2}C_{2} + O_{2}V_{2}C_{2}S_{2} - S_{2}S_{2}V_{2}O_{2} - C_{2}C_{2}O_{2}V_{2}\,\nonumber\\
\tau_{gg}&=& O_{2}V_{2}S_{2}C_{2} + V_{2}O_{2}C_{2}S_{2} - S_{2}S_{2}O_{2}V_{2} - C_{2}C_{2}V_{2}O_{2}\,\nonumber\\
\tau_{gh}&=& O_{2}O_{2}S_{2}S_{2} + V_{2}V_{2}C_{2}C_{2} - C_{2}S_{2}V_{2}V_{2} - S_{2}C_{2}O_{2}O_{2}\,\nonumber\\
\tau_{gf}&=& O_{2}O_{2}C_{2}C_{2} + V_{2}V_{2}S_{2}S_{2} - S_{2}C_{2}V_{2}V_{2} - C_{2}S_{2}O_{2}O_{2}\,\nonumber\\
\tau_{ho}&=& V_{2}S_{2}C_{2}O_{2} + O_{2}C_{2}S_{2}V_{2} - C_{2}O_{2}V_{2}C_{2} - S_{2}V_{2}O_{2}S_{2}\,\nonumber\\
\tau_{hg}&=& O_{2}C_{2}C_{2}O_{2} + V_{2}S_{2}S_{2}V_{2} - C_{2}O_{2}O_{2}S_{2} - S_{2}V_{2}V_{2}C_{2}\,\nonumber\\
\tau_{hh}&=& O_{2}S_{2}C_{2}V_{2} + V_{2}C_{2}S_{2}O_{2} - S_{2}O_{2}V_{2}S_{2} - C_{2}V_{2}O_{2}C_{2}\,\nonumber\\
\tau_{hf}&=& O_{2}S_{2}S_{2}O_{2} + V_{2}C_{2}C_{2}V_{2} - C_{2}V_{2}V_{2}S_{2} - S_{2}O_{2}O_{2}C_{2}\,\nonumber\\
\tau_{fo}&=& V_{2}S_{2}O_{2}C_{2} + O_{2}C_{2}V_{2}S_{2} - S_{2}V_{2}S_{2}O_{2} - C_{2}O_{2}C_{2}V_{2}\,\nonumber\\
\tau_{fg}&=& O_{2}C_{2}O_{2}C_{2} + V_{2}S_{2}V_{2}S_{2} - C_{2}O_{2}S_{2}O_{2} - S_{2}V_{2}C_{2}V_{2}\,\nonumber\\
\tau_{fh}&=& O_{2}S_{2}O_{2}S_{2} + V_{2}C_{2}V_{2}C_{2} - C_{2}V_{2}S_{2}V_{2} - S_{2}O_{2}C_{2}O_{2}\,\nonumber\\
\tau_{ff}&=& O_{2}S_{2}V_{2}C_{2} + V_{2}C_{2}O_{2}S_{2} - C_{2}V_{2}C_{2}O_{2} - S_{2}O_{2}S_{2}V_{2}\;
\ea
and nonsupersymmetric combinations \cite{cotrone}
\ba \label{a3}
\tau^{'}_{oo}&=& V_{2}O_{2}O_{2}O_{2} + O_{2}V_{2}V_{2}V_{2} - C_{2}S_{2}S_{2}C_{2} - S_{2}C_{2}C_{2}S_{2}\,\nonumber\\
\tau^{'}_{og}&=& O_{2}V_{2}O_{2}O_{2} + V_{2}O_{2}V_{2}V_{2} - C_{2}S_{2}C_{2}S_{2} - S_{2}C_{2}S_{2}C_{2}\,\nonumber\\
\tau^{'}_{oh}&=& O_{2}O_{2}O_{2}V_{2} + V_{2}V_{2}V_{2}O_{2} - S_{2}S_{2}S_{2}S_{2} - C_{2}C_{2}C_{2}C_{2}\,\nonumber\\
\tau^{'}_{of}&=& O_{2}O_{2}V_{2}O_{2} + V_{2}V_{2}O_{2}V_{2} - C_{2}C_{2}S_{2}S_{2} - S_{2}S_{2}C_{2}C_{2}\,\nonumber\\
\tau^{'}_{go}&=& O_{2}O_{2}S_{2}C_{2} + V_{2}V_{2}C_{2}S_{2} - S_{2}S_{2}V_{2}V_{2} - C_{2}C_{2}O_{2}O_{2}\,\nonumber\\
\tau^{'}_{gg}&=& O_{2}O_{2}C_{2}S_{2} + V_{2}V_{2}S_{2}C_{2} - S_{2}S_{2}O_{2}O_{2} - C_{2}C_{2}V_{2}V_{2}\,\nonumber\\
\tau^{'}_{gh}&=& V_{2}O_{2}S_{2}S_{2} + O_{2}V_{2}C_{2}C_{2} - S_{2}C_{2}O_{2}V_{2} - C_{2}S_{2}V_{2}O_{2}\,\nonumber\\
\tau^{'}_{gf}&=& O_{2}V_{2}S_{2}S_{2} + V_{2}O_{2}C_{2}C_{2} - C_{2}S_{2}O_{2}V_{2} - S_{2}C_{2}V_{2}O_{2}\,\nonumber\\
\tau^{'}_{ho}&=& V_{2}S_{2}C_{2}O_{2} + O_{2}C_{2}S_{2}V_{2} - S_{2}O_{2}V_{2}S_{2} - C_{2}V_{2}O_{2}C_{2}\,\nonumber\\
\tau^{'}_{hg}&=& O_{2}C_{2}C_{2}O_{2} + V_{2}S_{2}S_{2}V_{2} - C_{2}V_{2}V_{2}S_{2} - S_{2}O_{2}O_{2}C_{2}\,\nonumber\\
\tau^{'}_{hh}&=& O_{2}S_{2}C_{2}V_{2} + V_{2}C_{2}S_{2}O_{2} - C_{2}O_{2}V_{2}C_{2} - S_{2}V_{2}O_{2}S_{2}\,\nonumber\\
\tau^{'}_{hf}&=& O_{2}S_{2}S_{2}O_{2} + V_{2}C_{2}C_{2}V_{2} - C_{2}O_{2}O_{2}S_{2} - S_{2}V_{2}V_{2}C_{2}\,\nonumber\\
\tau^{'}_{fo}&=& O_{2}S_{2}O_{2}C_{2} + V_{2}C_{2}V_{2}S_{2} - S_{2}V_{2}S_{2}V_{2} - C_{2}O_{2}C_{2}O_{2}\,\nonumber\\
\tau^{'}_{fg}&=& O_{2}S_{2}V_{2}S_{2} + V_{2}C_{2}O_{2}C_{2} - C_{2}O_{2}S_{2}V_{2} - S_{2}V_{2}C_{2}O_{2}\,\nonumber\\
\tau^{'}_{fh}&=& V_{2}S_{2}O_{2}S_{2} + O_{2}C_{2}V_{2}C_{2} - S_{2}O_{2}C_{2}V_{2} - C_{2}V_{2}S_{2}O_{2}\,\nonumber\\
\tau^{'}_{ff}&=& O_{2}C_{2}O_{2}S_{2} + V_{2}S_{2}V_{2}C_{2} - S_{2}O_{2}S_{2}O_{2} - C_{2}V_{2}C_{2}V_{2}\;
\ea
\ba \label{a4}
\sigma_{oo}&=& O_{2}O_{2}O_{2}O_{2} + V_{2}V_{2}V_{2}V_{2} - C_{2}S_{2}S_{2}S_{2} - S_{2}C_{2}C_{2}C_{2}\,\nonumber\\
\sigma_{og}&=& O_{2}O_{2}V_{2}V_{2} + V_{2}V_{2}O_{2}O_{2} - S_{2}C_{2}S_{2}S_{2} - C_{2}S_{2}C_{2}C_{2}\,\nonumber\\
\sigma_{oh}&=& O_{2}V_{2}V_{2}O_{2} + V_{2}O_{2}O_{2}V_{2} - S_{2}S_{2}S_{2}C_{2} - C_{2}C_{2}C_{2}S_{2}\,\nonumber\\
\sigma_{of}&=& O_{2}V_{2}O_{2}V_{2} + V_{2}O_{2}V_{2}O_{2} - C_{2}C_{2}S_{2}C_{2} - S_{2}S_{2}C_{2}S_{2}\,\nonumber\\
\sigma_{ho}&=& O_{2}S_{2}C_{2}O_{2} + V_{2}C_{2}S_{2}V_{2} - S_{2}O_{2}V_{2}C_{2} - C_{2}V_{2}O_{2}S_{2}\,\nonumber\\
\sigma_{hg}&=& O_{2}S_{2}S_{2}V_{2} + V_{2}C_{2}C_{2}O_{2} - S_{2}O_{2}O_{2}S_{2} - C_{2}V_{2}V_{2}C_{2}\,\nonumber\\
\sigma_{hh}&=& O_{2}C_{2}S_{2}O_{2} + V_{2}S_{2}C_{2}V_{2} - S_{2}V_{2}O_{2}C_{2} - C_{2}O_{2}V_{2}S_{2}\,\nonumber\\
\sigma_{hf}&=& O_{2}C_{2}C_{2}V_{2} + V_{2}S_{2}S_{2}O_{2} - C_{2}O_{2}O_{2}C_{2} - S_{2}V_{2}V_{2}S_{2}~.
\ea
We use the notation ($i=o,g,h,f$):
\ba \label{a5}
T_{io}&=& \tau_{io}+\tau_{ig}+\tau_{ih}+\tau_{if}~,\qquad T_{ig}= \tau_{io}+\tau_{ig}-\tau_{ih}-\tau_{if}~,\nonumber \\
T_{ih}&=& \tau_{io}-\tau_{ig}+\tau_{ih}-\tau_{if}~,\qquad T_{if}= \tau_{io}-\tau_{ig}-\tau_{ih}+\tau_{if}~,
\ea
and likewise for the $\tau^{'}$'s ($T^{'}_{ij}$) and the $\sigma$'s
($S_{ij}$); a superscript ``F'' or ``B'' for the T's will denote the
Fermionic or Bosonic part of the characters. In our convention, 
$T_{ij}^{B \pm F} = T_{ij}^B \pm T_{ij}^F$, etc. We also used the definitions
\ba
\Lambda_{m,2n}&=&\sum_{m,n}\frac{1+(-1)^{n}}{2}\Lambda_{m,n}~, \qquad
\quad \Lambda_{m,2n+1}=\sum_{m,n}\frac{1-(-1)^{n}}{2} \Lambda_{m,n}~, \nonumber\\
\Lambda_{m+1/2,2n} &=&\sum_{m,n}\frac{1+(-1)^{n}}{2}
\Lambda_{m+\frac{1}{2},n} \quad ,  \quad
\Lambda_{m+1/2,2n+1} =\sum_{m,n}\frac{1-(-1)^{n}}{2}
\Lambda_{m+\frac{1}{2},n}~, \label{a6}
\ea
where $\Lambda_{m,n}$ denotes the standard $(1,1)$ momentum and winding lattice 
and
\be
P_{m+a} (\tau) \equiv \sum_m q^{\pi \alpha' (m+a)^2 \over R^2}
\quad , \quad
W_{n+b} (\tau) \equiv \sum_n q^{{\pi \over 4 \alpha'} {(n+b)^2
R^2}} \ . \label{a7} 
\ee


\end{document}